\documentclass[lettersize,journal]{IEEEtran}
\usepackage{amsmath,amsfonts,algorithmic,algorithm,array,bm,color,cite,epstopdf,graphicx,mathtools,multirow,pdfpages,soul,textcomp,xcolor,stfloats,url,verbatim,}
\usepackage{graphicx}
\usepackage[colorlinks=true, allcolors=blue]{hyperref}
\DeclarePairedDelimiter\ceil{\lceil}{\rceil}

\begin{document}
\title{A Learning Convolutional Neural Network Approach for Network Robustness Prediction}

\author{Yang~Lou,~
	Ruizi~Wu, Junli~Li,
	Lin~Wang,~
	Xiang~Li,~
	and
	Guanrong~Chen~
	\thanks{Yang Lou is with the Department of Computing and Decision Sciences, Lingnan University, Hong Kong, China, and also with the Key Laboratory of System Control and Information Processing, Ministry of Education of China, Shanghai 200240, China (e-mail: felix.lou@ieee.org).}
	\thanks{Ruizi~Wu and Junli~Li are with the College of Computer Science, Sichuan Normal University, Chengdu 610066, China (e-mail: vridge@foxmail.com; lijunli@sicnu.edu.cn).}%
	\thanks{Lin Wang is with the Department of Automation, Shanghai Jiao Tong University, Shanghai 200240, China, and also with the Key Laboratory of System Control and Information Processing, Ministry of Education, Shanghai 200240, China (e-mail: wanglin@sjtu.edu.cn).}%
	\thanks{Xiang Li is with the Institute of Complex Networks and Intelligent Systems, Shanghai Research Institute for Intelligent Autonomous Systems, Tongji University, Shanghai 201210, and also with the Department of Control Science and Engineering, Tongji University, Shanghai 200240, China (e-mail: lix2021@tongji.edu.cn).}  
	\thanks{Guanrong Chen is with the Department of Electrical Engineering, City University of Hong Kong, Hong Kong, China (e-mail:eegchen@cityu.edu.hk).}%
	
	\thanks{(\textit{Yang Lou and Ruizi Wu contributed equally to this work})}
	\thanks{(\textit{Corresponding author: Yang Lou and Lin Wang})}
}
\maketitle
\begin{abstract}
	Network robustness is critical for various societal and industrial networks again malicious attacks. In particular, connectivity robustness and controllability robustness reflect how well a networked system can maintain its connectedness and controllability against destructive attacks, which can be quantified by a sequence of values that record the remaining connectivity and controllability of the network after a sequence of node- or edge-removal attacks. Traditionally, robustness is determined by attack simulations, which are computationally very time-consuming or even practically infeasible. In this paper, an improved method for network robustness prediction is developed based on learning feature representation using convolutional neural network (LFR-CNN). In this scheme, higher-dimensional network data are compressed to lower-dimensional representations, and then passed to a CNN to perform robustness prediction. Extensive experimental studies on both synthetic and real-world networks, both directed and undirected, demonstrate that 1) the proposed LFR-CNN performs better than other two state-of-the-art prediction methods, with significantly lower prediction errors; 2) LFR-CNN is insensitive to the variation of the network size, which significantly extends its applicability; 3) although LFR-CNN needs more time to perform feature learning, it can achieve accurate prediction faster than attack simulations; 4) LFR-CNN not only can accurately predict network robustness, but also provides a good indicator for connectivity robustness, better than the classical spectral measures.
\end{abstract}

\section{Introduction}
\IEEEPARstart{A}{complex} network is a graph consisting of large numbers of nodes and edges with complicated connections. Many natural and engineering systems can be modeled as complex networks, and then studied using graph theory and network analysis tools. The study of complex networks attracts increasing interest from research communities in various scientific and technological fields, including computer science, systems engineering, applied mathematics, statistical physics, biological sciences, and social sciences \cite{Barabasi2016NS,Newman2010N,Chen2014Book,Chen2019Book}.

In the pursuit of networked systems control for beneficial applications, the \textit{network controllability} \cite{Liu2011N,Yuan2013NC,Posfai2013SR,Menichetti2014PRL,Pan2014PS,Motter15CHAOS,Wang2016AUTO,Hou2016TCASI,Liu2016RMP,Wang2017RSPTA,Wang2017SR,Hou2017TCNS,Zhang2017TAC,Xiang2019CSM,Wu2020TCASI,Hou2020TAC} is a fundamental issue, which refers to the ability of a network of interconnected dynamic systems in changing from any initial state to any desired state under feasible control input within finite time \cite{Xiang2019CSM}. The \textit{network connectivity} is fundamentally important for a network to function, affecting particularly the network controllability \cite{Xiang2019CSM} and synchronizability \cite{Shi2013CSM}. It is easy to see that good controllability requires good connectivity, but good connectivity does not necessarily guarantee good controllability \cite{Lou2021TNSE}. In fact, network connectivity and controllability have very different characteristics and measures: the former is guaranteed by a sufficient number of edges, while the later further requires a proper organization of the sufficient number of edges.

Today, malicious attacks and random failures widely exist in many engineering and technological facilities and processes, which degrade or even destroy certain network functions typically through destructing the network connectivity. Therefore, it is essential to strengthen the network connectivity against such attacks and failures \cite{Holme2002PRE,Shargel2003PRL,Schneider2011PNAS,Bashan2013NP,Fan2020NMI,Wang2021TEVC,Lou2021TNSE,Grassia2021NC}. In general, destructive attacks and failures take place in the forms of node- and edge-removals, which may cause significant degeneration of network connectivity and controllability. In such situations, the abilities of a network to maintain its connectivity and controllability against attacks or failures are of great concerns, which are referred to as the \textit{connectivity robustness} and \textit{controllability robustness}, respectively.

Connectivity robustness is commonly measured by using the change of the portion of nodes in the largest connected component (LCC) \cite{Schneider2011PNAS} that survives from a sequence of attacks. A network is deemed more robust against attacks if it can always maintain higher values of the fractions of LCC nodes throughout an attack process. The investigation and optimization of connectivity robustness using this measure emphasize on protecting the LCC. Given certain practical constraints, e.g., node degree preservation, connectivity robustness can be enhanced by edge rewiring, which actually imposes disturbances onto the network structure \cite{Wu2011PRE,Zeng2012PRE,Louzada2013JCN,Schneider2013SR,Liang2015CPL,Chan2016DMKD,Lou2021TCASII,Wang2020TEVC,Wang2021TEVC}. After some edge rewiring operations, whether such disturbance enhances the robustness or not has to be evaluated, typically by using very time-consuming attack simulations. As a remedy, several easy-to-access indicators, e.g. assortativity \cite{Newman2003PRE} and spectral measures \cite{Perra2008PRE}, are adopted for estimating the network connectivity robustness. For example, it is found that onion-like structured heterogeneous networks with positive assortativity coefficients are robust against attacks \cite{Schneider2011PNAS,Wu2011PRE,Tanizawa2012PRE,Hayashi2018SR}. However, these measures have limited scopes of applications, and therefore the time-consuming attack simulation remains as the main approach today.

Controllability robustness is generally measured using the change of density of driver nodes, at which external control signals can be imposed as input. A network is deemed more robust against attacks, if it can maintain a lower density of driver nodes throughout an attack process. The studies and optimization of controllability robustness using this measure emphasize on maintaining a low demand of additional driver nodes. Although controllability robustness can be enhanced by edge rewiring as in connectivity robustness enhancement, their objective functions in optimization are very different. In fact, on top of the connectedness, the way the nodes are connected makes a huge impact on the controllability. For example, it is observed that a power-law degree distribution does not necessarily imply weak controllability robustness; while multi-chain \cite{Yan2016SR} and multi-loop \cite{Lou2018TCASI,Chen2019TCASII} structures significantly strengthen the controllability robustness. It is empirically found that extreme homogeneity is necessary for the optimal topology that has the best controllability robustness against random node attacks \cite{Lou2020TCASI}. Likewise, attack simulation is a main approach to measuring network controllability robustness today, which however is even more time-consuming than measuring the network connectivity discussed above.

For both connectivity and controllability robustness enhancements, deep neural networks\cite{Iiduka2021TCYB,Xiao2021TCYB,Sun2021TCYB} provide a useful tool for computation, optimization and analysis. Successful deep learning applications on complex networks include network robustness prediction using convolutional neural networks (CNNs) \cite{Schmidhuber2015NN,Lou2020TCYB,Dhiman2020MLN,Lou2021TNNLS,Lou2021TNSE}, and critical node identification using deep reinforcement learning \cite{Fan2020NMI} and graph attention networks \cite{Grassia2021NC}. Main advantages of CNN-based approaches for robustness prediction include: 1) the method is straightforward, where the adjacency matrix of a complex network is treated as a gray-scale image, and then the classification (if any) and regression tasks are same as in image processing. 2) The performance of CNN-based approach is stable and reliable: all types of network adjacency matrices are acceptable as input, which is also shift-invariant \cite{Zhang2019ICML}, namely shuffling and transposing pixels of an image (while keeping the network topology unchanged) does not degrade the performance of the prediction \cite{Lou2021TNNLS,Lou2021TNSE}. In addition, it has been experimentally demonstrated that CNN is tolerable to slightly changes of the network size.

However, the above CNN-based approaches cannot guarantee the prediction performance when the input size has significant changes (e.g., $\pm20\%$ or more) from the training samples. In addition, since many complex networks are sparse, the gray-scale images converted from network adjacency matrices typically contain a large amount of useless information, where quite a lot of pixels can be removed or compressed.

To overcome the aforementioned issues, a learning feature representation-based CNN (LFR-CNN) approach is proposed in this paper for precise network robustness prediction. LFR-CNN consists of an LFR module and a CNN. The LFR module performs feature extraction and dimensionality reduction, so that the size of input to the CNN for prediction can be significantly reduced, and simultaneously redundant information can be filtered out.

The following text is organized as follows. Section \ref{sec:pre} reviews the measures of network connectivity and controllability robustness against destructive node-removal attacks. Section \ref{sec:cnn} introduces the details of the proposed LRF-CNN. Section \ref{sec:exp} presents experimental results with analysis and comparison. Section \ref{sec:end} concludes the investigation.

\section{Robustness of Complex Networks} \label{sec:pre}

The concepts and calculations of connectivity robustness and controllability robustness are introduced in this section, where connectivity robustness reflects how well a networked system can maintain its connectedness under a sequence of node-removal attacks, while controllability robustness reflects how well it can maintain its controllable state. In this paper, only node-removal attacks are investigated, while edge-removal attacks can be studied in a similar manner.

\subsection{Connectivity Robustness}\label{sub:nec}

An undirected network is connected if and only if for each pair of nodes there is a path between them. A directed network is \textit{weakly connected} if it remains to be connected after all the directions are removed. Both \textit{connectedness} and \textit{weak connectedness} are employed as measures of the network connectivity in this paper, for undirected and directed networks respectively.

Under a sequence of node-removal attacks, connectivity robustness is evaluated using the fraction of nodes in LCC after each node-removal \cite{Schneider2011PNAS}, as follows:
\begin{equation}\label{eq:nlc}
	p(i)=\frac{N_{\text{LCC}}(i)}{N-i}\,,~~i=0,1,\ldots,N-1\,,
\end{equation}
where $p(i)$ is the fractions of nodes in LCC after a total number of $i$ nodes removed; $N_{\text{LCC}}(i)$ is the number of nodes in LCC after a total number of $i$ nodes have been removed from the network; $N$ is the number of nodes in the network before being attacked. When these values are plotted versus the fraction of removed nodes, a curve is obtained, called the \textit{connectivity curve}.

\subsection{Controllability Robustness}

For a linear time-invariant networked system $\dot{{\bf x}}=A{\bf x}$+$B{\bf u}$, where $A$ and $B$ are constant matrices of compatible dimensions, and $\bf x$ and $\bf u$ are the state vector and control input, respectively. The system is \textit{state controllable} if and only if the controllability matrix $[B\ AB\ A^2B\ \cdots A^{N-1}B]$ has a full row-rank, where $N$ is the dimension of $A$, which is also the size of the network in the present study. It is shown \cite{Liu2011N} that, for a directed network, identifying the set of the minimum number of driver nodes $N_D$ can be converted to searching for a maximum matching of the network: $N_D=\text{max}\{1, N-|E^*|\}$, where $|E^*|$ is the number of edges in the maximum matching $E^*$. For an undirected network, the minimum number of needed driver nodes can be calculated using the exact controllability formula \cite{Yuan2013NC}: $N_D=\text{max}\{1, N-\text{rank}(A)\}$. Then, the network controllability robustness is calculated as follows:
\begin{equation}\label{eq:ndi}
	q(i)=\frac{N_D(i)}{N-i}\,,\ \ i=0,1,\ldots,N-1,
\end{equation}
where $N_D(i)$ is the number of driver nodes needed to retain the network controllability after a total of $i$ nodes have been removed, and $N$ is the network size. When these values are plotted versus the fraction of removed nodes, a curve is obtained, called the \textit{controllability curve}.

\subsection{Error Measures}

For either connectivity or controllability, consider three curves: $\textbf{s}_t=[s_t(0),\cdots,s_t(N-1)]$ denotes the true curve obtained by attack simulations, and $\textbf{s}_1=[s_1(0),\cdots,s_1(N-1)]$ and $\textbf{s}_2=[s_2(0),\cdots,s_2(N-1)]$ denote two predicted curves, respectively. The difference between the true curve and a predicted curve is calculated by $\bm{\xi}_{\alpha}=|\textbf{s}_t-\textbf{s}_{\alpha}|$, where $\bm{\xi}_{\alpha}=[\xi_{\alpha}(0),\cdots,\xi_{\alpha}(N-1)]$ is the sequence of errors between the two curves, where $\xi_{\alpha}(i)=|s_t(i)-s_{\alpha}(i)|$, for $\alpha=1~\text{or}~2$, and $i=0,1,\cdots,N-1$.

The \textit{prediction error} $\bar{\xi}_{\alpha}$ is then calculated by
\begin{equation}\label{eq:avg_xi}
	\bar{\xi}_{\alpha}=\frac{1}{N}\sum_{i=0}^{N-1}{\xi(i)}_{\alpha}\,.
\end{equation}
The vector $\bm{\xi}_{\alpha}$ can be used to visualize the prediction errors throughout the attack process. The scalar $\bar{\xi}_{\alpha}$ measures the \textit{overall} prediction error, i.e., $\bar{\xi}_1<\bar{\xi}_2$ means that the predicted curve $\textbf{s}_1$ obtains lower prediction error than $\textbf{s}_2$.

For notational convenience, the integer index sequence $i=0,1,\ldots,N-1$, will be replaced by the fractional index sequence $\delta=0,\frac{1}{N},\ldots,\frac{N-1}{N}$, thereby equivalently replacing $n_D(i)$, with $n_D(\delta)$.

\section{Performance Predictor}\label{sec:cnn}

This section briefly reviews the predictor for controllability robustness (PCR) \cite{Lou2020TCYB}, which employs a VGG-structured CNN \cite{Simonyan2014arXiv} and PATCHY-SAN \cite{Niepert2016ICML} consisting of an LFR-based 1D-CNN. Pros and cons of these two approaches are discussed. Then, a structural LFR-CNN is designed by incorporating the LFR module and a simplified VGG-structured CNN. LFR-CNN has a parameter magnitude significantly greater than PATCHY-SAN, but less than PCR.

\subsection{Convolutional Neural Network}

\begin{figure}[htbp]
	\centering \includegraphics[width=\linewidth]{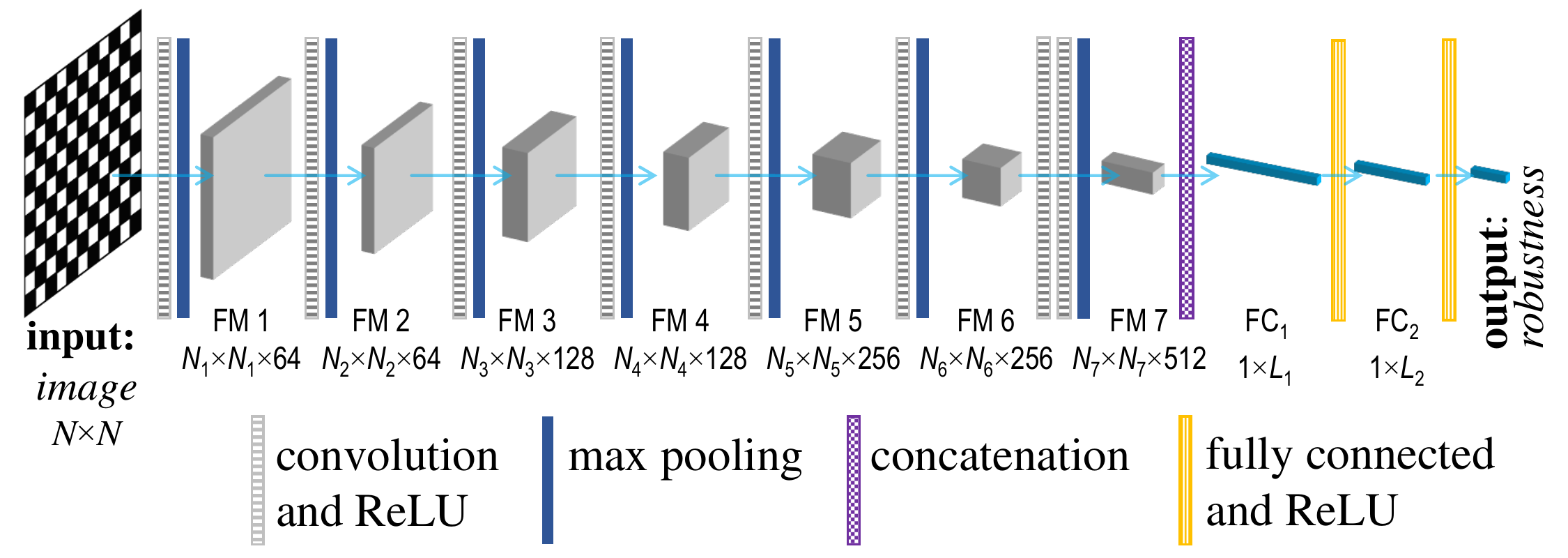}
	\caption{CNN structure of PCR. The input is adjacency matrix; the output is an $N$-vector. For $N=1000$, seven feature map (FM) groups are installed with $N_i=\ceil{N/2^{(i+1)}}$, for $i=1,2,\ldots,7$. The concatenation layer reshapes the matrix to a vector, from FM 7 to FC 1, i.e., FC1=$N_7\times N_7\times 512$ and FC2=$4096$ \cite{Lou2020TCYB}. }\label{fig:cnn}
\end{figure}

PCR is a CNN-based framework for predicting the controllability robustness \cite{Lou2020TCYB}, which has also been applied to predict connectivity robustness \cite{Lou2021TNSE}. The CNN structure of the PCR is shown in Fig. \ref{fig:cnn}. Network adjacency matrices are converted to gray-scale images and then used directly as input to CNN. Both classification and regression tasks can be performed using such an image-processing mechanism. Due to a sufficiently large source of training data that can be generated using various synthetic network models, tens of millions of internal parameters in a CNN can be properly trained.

The mean-squared error between the predicted connectivity or controllability curve $\hat{v}$ and true curve $v$ is used as the loss function:
\begin{equation}\label{eq:lf}
	\mathcal{L} = \frac{1}{N+1} \sum_{i=0}^{N}||\hat{v}(i)-v(i)||\,,
\end{equation}
where $\hat{v}(i)$ represents the predicted connectivity or controllability value when a total proportion of $i/N$ nodes have been removed from the network; $v(i)$ represents the corresponding true value obtained by attack simulation; $||\cdot||$ is the Euclidean norm. The training process aims at adjusting the internal parameters\cite{Iiduka2021TCYB}, with the objective of minimizing $\mathcal{L}$.

\subsection{PATCHY-SAN}

Complex network data have distinguished continuous and discrete attributes that are different from general image data. A group of recurrent neural networks, namely the graph neural networks (GNNs) \cite{Kipf2016arXiv,Hamilton2017NIPS,Hamilton2020Book}, are specifically designed for processing graph data. Specifically, lower-dimensional representations are generated from compacting higher-dimensional raw graph data, and then classification or/and regression tasks are performed by processing the lower-dimensional representation data. PATCHY-SAN \cite{Niepert2016ICML}, as a successful GNN technique, processes graph data with \textit{selecting}, \textit{assembling}, and \textit{normalizing} (SAN) operations, detailed below.

\subsubsection{Node Sequence Selection}

A fixed-length sequence of $W$ nodes are selected from the $N$ nodes in the network. Nodes are arranged in descending order according to certain importance measure. Thus, for different networks, similar important nodes are arranged in similar ranks in the node sequence.

Node sequence selection is the process of sorting and identifying critical nodes. Each node is assigned a score via a labeling procedure, where node centrality measures such as degree and betweenness are used to describe the importance of a node. Then, all the nodes are sorted in descending order of the labeling scores; the first $W$ nodes are selected as the node sequence. A receptive fields of size $g$ will be created for each node in the selected sequence. Each receptive field is constructed by \textit{assembling} and \textit{normalizing} as introduced in the following. Note that if $N<W$, all-zero receptive fields are added for padding.

\subsubsection{Neighborhood Assembly}

A set of neighboring nodes is collected for each node in the selected sequence. A breadth-first search is used to collect the neighborhood field, namely if there is not enough neighboring nodes collected in the current depth, then search in the one-step further neighborhoods, and so on, until at lease $g$ neighboring nodes are collected, or no more neighboring node to explore.

\subsubsection{Normalization}

The extracted neighborhood data are ranked to create the normalized receptive fields. The normalization process also imposes an order on the neighboring field for each selected node such that the unordered neighboring field is mapped into an embedding vector space in a linear order. The orders of nodes are determined by a labeling procedure using node centrality measures. In the resultant normalized vector, the root node is assigned as the first element, followed by the second to the $g$-th neighboring nodes. This normalization procedure leverages graph labeling on the neighboring nodes of the root node.

To this end, an $N$-node network is represented by a $W$-unit receptive field, where each receptive field is a $g\times h$ matrix, with $h$ representing the number of attributes used for the neighboring nodes. Since generally $W\leq N$, $g\ll N$, and $h\ll N$, an $N^2$ adjacency matrix is mapped to a compressed representation of size $Wgh$, which will be reshaped and then passed to a 1D-CNN for further processing in PATCHY-SAN.

Since this procedure generates learned feature representations for graph data, it is named an LFR module.

\subsection{LFR-CNN}

PCR is straightforward and fast, while PATCHY-SAN extracts topological features first. The input of PCR is the raw adjacency matrix. Since many real-world networks are sparse, which have much fewer edges than the possible maximum number of edges, the input adjacency matrix contains a lot of meaningless information that can be removed or compressed. In contrast, PATCHY-SAN employs a shallow 1D-CNN structure. Empirically, if properly trained and used, deeper neural networks with more layers and parameters are prone to having better performances than those with fewer layers and parameters, especially for large-scale complex network data.

\begin{table}[htbp]
	\centering \caption{Comparison of PCR, PATCHY-SAN, LFR-CNN in terms of representation, representation size, number of layers, and magnitude of number of parameters.}
	\begin{tabular}{|l|l|c|c|c|}\hline
		& \multicolumn{1}{c|}{\begin{tabular}[c]{@{}c@{}}Converted\\Representation\end{tabular} } & Size & \begin{tabular}[c]{@{}c@{}}Feature\\Maps\end{tabular} & Parameters \\ \hline
		PCR & \begin{tabular}[c]{@{}l@{}}Gray-Scale\\Image\end{tabular} & $N^2$ & $7(6)$ & $2.4\times10^7$ \\ \hline
		PATCHY-SAN & \begin{tabular}[c]{@{}l@{}}LFR \end{tabular} & $Wgh$ & $2$ & $5.1\times10^5$ \\ \hline
		LFR-CNN & \begin{tabular}[c]{@{}l@{}}LFR \end{tabular} & $Wgh$ & $3$ & $6.0\times10^6$ \\ \hline
	\end{tabular}\label{tab:cmp}
\end{table}

Table \ref{tab:cmp} shows that PCR converts an $N^2$ adjacency matrix to an gray-scale image without compression, while for PATCHY-SAN an adjacency matrix is compressed to an LFR of size $Wgh$. The core components of PCR and PATCHY-SAN are a 2D-CNN and a 1D-CNN, respectively. A CNN with 7 feature map (FM) groups (or 6-FM for small-sized networks) in PCR requires training a total number of $2.4\times10^7$ internal parameters, while the 1D-CNN in PATCHY-SAN requires training $5.1\times10^5$ parameters.

In this paper, an LFR-CNN is proposed by installing a 2D-CNN (similar to PCR, but with shallower structure) following the LFR module of PATCHY-SAN. Compared to PCR and PATCHY-SAN, LFR-CNN has the following advantages: 1) a 2D-CNN can be more powerful than the 1D-CNN in PATCHY-SAN. 2) With LFR, the required number of FMs in 2D-CNN can be significantly reduced, and more importantly the required number of FMs does not need to change for different network sizes. 3) LFR-CNN requires an intermediate number of training parameters, i.e., $6.0\times10^6$. LFR-CNN achieves a balance in CNN structure and magnitude of number of parameters between PCR and PATCHY-SAN.

\begin{figure}[htbp]
	\centering
	\includegraphics[width=\linewidth]{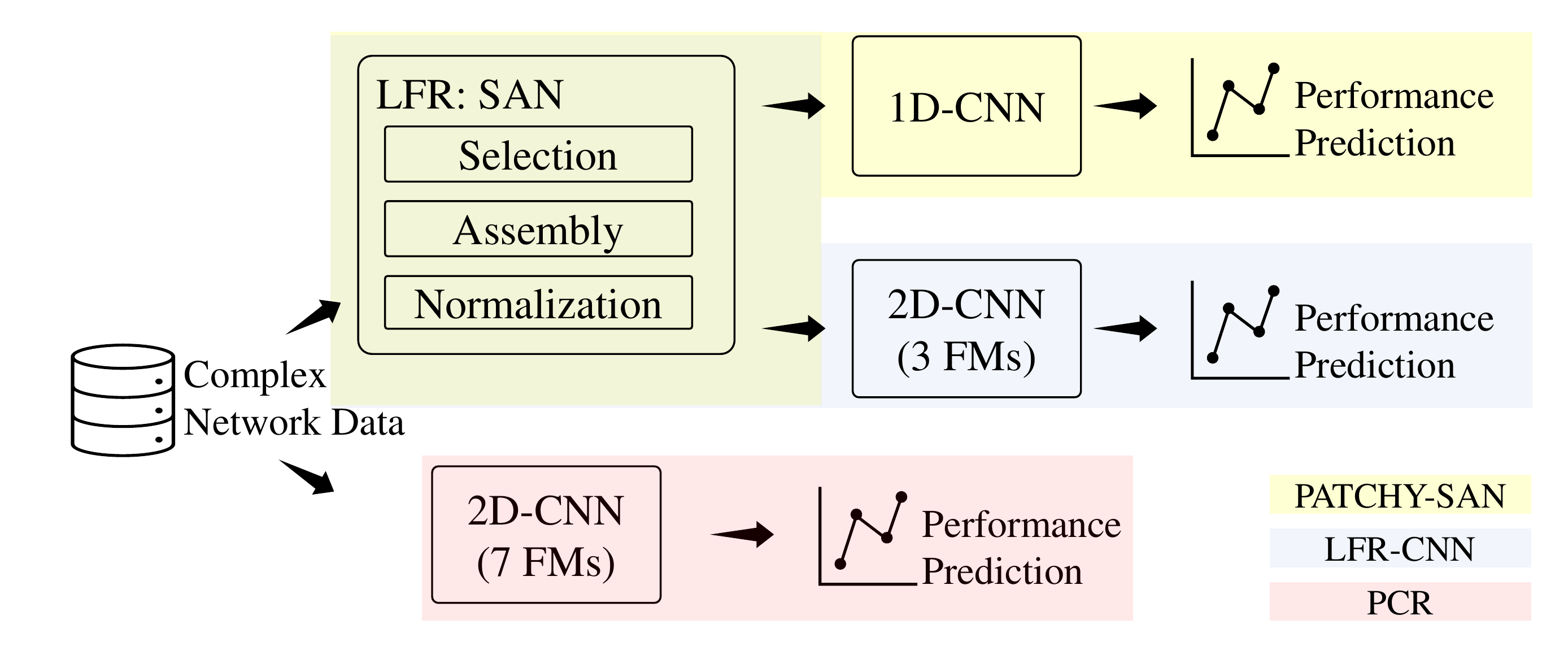}\\
	\caption{General framework of PATCHY-SAN, LFR-CNN, and PCR: PATCHY-SAN and LFR-CNN share the common module of LFR performing the selection, assembly, and normalization (SAN) tasks. LFR-CNN and PCR share a similar VGG-structured 2D-CNN module.}\label{fig:gen_flow}
\end{figure}

The different structures of PCR, PATCHY-SAN, and LFR-CNN are shown in Fig. \ref{fig:gen_flow}, where the LFR module consists of selection, assembly, and normalization operations. Given the same LFR as the input, a 2D-CNN can capture more feature details than a 1D-CNN, therefore is more suitable to be applied to large-scale complex network data. The proposed LFR-CNN naturally combines PATCHY-SAN and PRC by incorporating their advantages.

Similarly to PCR, a VGG-structured \cite{Simonyan2014arXiv} CNN is installed in LFR-CNN. For network sizes around $N=1000$, PCR needs seven FM groups to perform prediction. When the network size is reduced (e.g., $N=500$), the number of FMs can be reduced (e.g., 6 FMs). In contrast, since raw graph data are compressed by the LFR module, the CNN in LFR-CNN is not necessary to be adjusted if the network sizes are not significantly changed. Specifically, as shown in the experimental studies, LFR-CNN is able to process different network sizes $N\in[350,1300]$ using the same 3-FM CNN.

\begin{figure}[htbp]
	\centering
	\includegraphics[width=\linewidth]{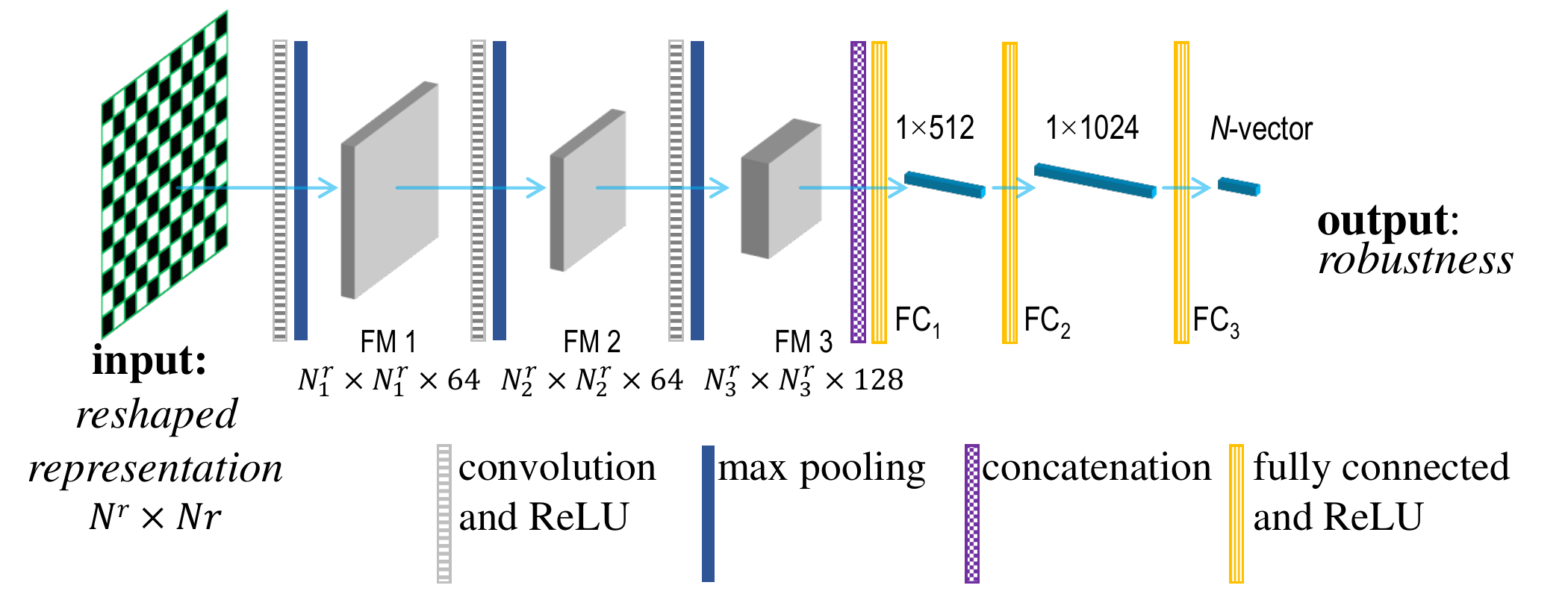}
	\caption{The simplified 2D-CNN structure with three feature map groups installed with $N^{r}_{i}=\ceil{N^{r}/2^{(i+1)}}$, for $i=1,2,3$, where $N^{r}\times N^{r}$ is the size of the input reshaped representation. The concatenation layer reshapes the matrix to a vector from FM 3 to FC 1. }\label{fig:cnn31}
\end{figure}
\begin{table}[htbp]
	\centering \caption{Parameter setting of the 3-FM 2D-CNN installed in LFR-CNN.}
	\begin{tabular}{|c|c|c|c|c|} \hline
		Group & Layer & Kernel Size & Stride & \begin{tabular}[c]{@{}c@{}}Output\\ Channel\end{tabular} \\ \hline
		\multirow{2}{*}{Group 1} & Conv7-64 & $7\times7$ & 1 & 64 \\ \cline{2-5}
		& Max2 & $2\times2$ & 2 & 64 \\ \hline
		\multirow{2}{*}{Group 2} & Conv5-64 & $5\times5$ & 1 & 64 \\ \cline{2-5}
		& Max2 & $2\times2$ & 2 & 64 \\ \hline
		\multirow{2}{*}{Group 3} & Conv3-128 & $3\times3$ & 1 & 128 \\ \cline{2-5}
		& Max2 & $2\times2$ & 2 & 128 \\ \hline
	\end{tabular}\label{tab:cnnpara}
\end{table}
The detailed structure is illustrated in Fig. \ref{fig:cnn31}, and the parameters are summarized in Table \ref{tab:cnnpara}. Each group of FM1--FM3 contains a convolution layer, a ReLU performing the activation function $f(x)=\text{max}(0,x)$ \cite{Glorot2011ICAIS}, and a max pooling layer. The output of each hidden layer is summed up, rectified by a ReLU, and then transmitted to the next layer. To that end, max pooling layers will reduce the data dimension as input to the next layer. Then, two fully-connected layers are installed to map feature representations and reshape the regression output. The same loss function as in PCR is employed, as shown in Eq. (\ref{eq:lf}).

\section{Experimental Studies}\label{sec:exp}

A total of 9 synthetic network models are simulated, including the Erd{\"{o}}s-R{\'e}nyi (ER) random-graph \cite{Erdos1964RG}, Barab{\'a}si--Albert (BA) scale-free \cite{Barabasi1999SCI,Barabasi2009SCI}, generic scale-free (SF) \cite{Goh2001PRL}, onion-like generic scale-free (OS) \cite{Schneider2011PNAS}, Newman--Watts small-world (SW-NW) \cite{Newman1999PLA}, Watts--Strogatz small-world (SW-WS) \cite{Watts1998N}, \textit{q}-snapback (QS) \cite{Lou2018TCASI}, random triangle (RT) \cite{Chen2019TCASII} and random hexagon (RH) \cite{Chen2019TCASII} 
networks.

Specifically, a BA network is generated according to the preferential attachment scheme \cite{Barabasi1999SCI}, while an SF network is generated according to a set of predefined weights $w_{i}=(i+\theta)^{-\sigma}$, where $i=1,2,\ldots,N$, $\sigma\in[0,1)$ and $\theta\ll N$. Two nodes $i$ and $j$ are randomly picked with a probability proportional to their weights $w_i$ and $w_j$, respectively. An OS network is generated based on an SF, with $2N$ rewiring operations towards assortativity maximization. The degree distributions of BA, SF, and OS all follow the power law.

Both SW-NW and SW-WS start from an $N$-node loop having $K$ ($=2$) connected nearest-neighbors. The difference is that additional edges are added without removing any existing edges in SW-NW \cite{Newman1999PLA}; while rewiring operations are performed in SW-WS \cite{Watts1998N}.

QS consists of a backbone chain and multiple snapback edges \cite{Lou2018TCASI}. RT and RH consist of random triangles and hexagons, respectively \cite{Chen2019TCASII}.

To exactly control the number of generated edges to be $M$, uniformly-randomly adding or removing edges can be performed. A directed network can be converted to an undirected network by removing the direction. However, when converting an undirected network to be directed, it should follow some specific patterns, e.g., a directed backbone chain in QS and a directed loop in SW-NW and SW-WS should be ensured; while for some other edges, directions can be assigned randomly.

For each synthetic network, $1000$ instances are randomly generated for training, thus there are $1000\times9=9000$ training samples in total. In addition, two different sets of $100\times9=900$ samples are used for cross validation and testing, respectively.

The size of each synthetic network is randomly determined in three different settings, namely, 1) set $N\in[350,650]$ (with an average $\bar{N}=500.5$) for the experiments of predicting connectivity and controllability robustness in Subsections \ref{sub:exp_dir}, \ref{sub:exp_real}, \ref{sub:exp_und}, \ref{sub:exp_ft}, \ref{sub:exp_rt}, and \ref{sub:exp_spm}; 2) set $N\in[700,1300]$ (with an average $\bar{N}=999.8$) for the scalability investigation in Subsection \ref{sub:exp_size}; 3) set $N\in[700,900]$ (with an average $\bar{N}=800.0$) for the study of the influence of information loss on the three comparative approaches in Subsection \ref{sub:exp_pcr}.

The average degrees are also assigned randomly. The ranges are set differently for various network models. For SW $\langle k\rangle\in[2.5,5]$, for RH, $\langle k\rangle\in[2,4]$, for RT, $\langle k\rangle\in[1.5,3]$, while for other models, $\langle k\rangle\in[3,6]$. The overall average degree of the training network is $4.33$, while that of the testing network is $4.36$, with data obtained by performing posterior statistics.

The proposed LFR-CNN is compared with PATCHY-SAN \cite{Niepert2016ICML} and PCR \cite{Lou2020TCYB,Lou2021TNSE} in predicting the connectivity and controllability robustness for both synthetic and real-world networks under various node-removal attacks, including random attack (RA), targeted betweenness-based (TB) attack, and targeted degree-based (TD) attack. For PCR, a 6-FM CNN is used for $N<700$ and a 7-FM structure is used for $N\geq700$. For PATCHY-SAN and LFR-CNN, the structures remain the same for all networks with $N\in[350,1300]$. For LFR, set the length of the selected node sequence to be $W=500$ for $N<700$, and $W=1000$ for $N\geq700$; the receptive field size $g=10$; the number of attributes $h=2$ (the two default attributes are node degree and clustering coefficient).

All experiments are performed on a PC Intel (R) Core i7-8750H CPU @ 2.20GHz, which has memory (RAM) 16 GB with running Windows 10 Home 64-bit Operating System.

\subsection{Predicting Controllability Robustness for Directed Networks} \label{sub:exp_dir}
\begin{figure}[htbp]
	\centering\includegraphics[width=\linewidth]{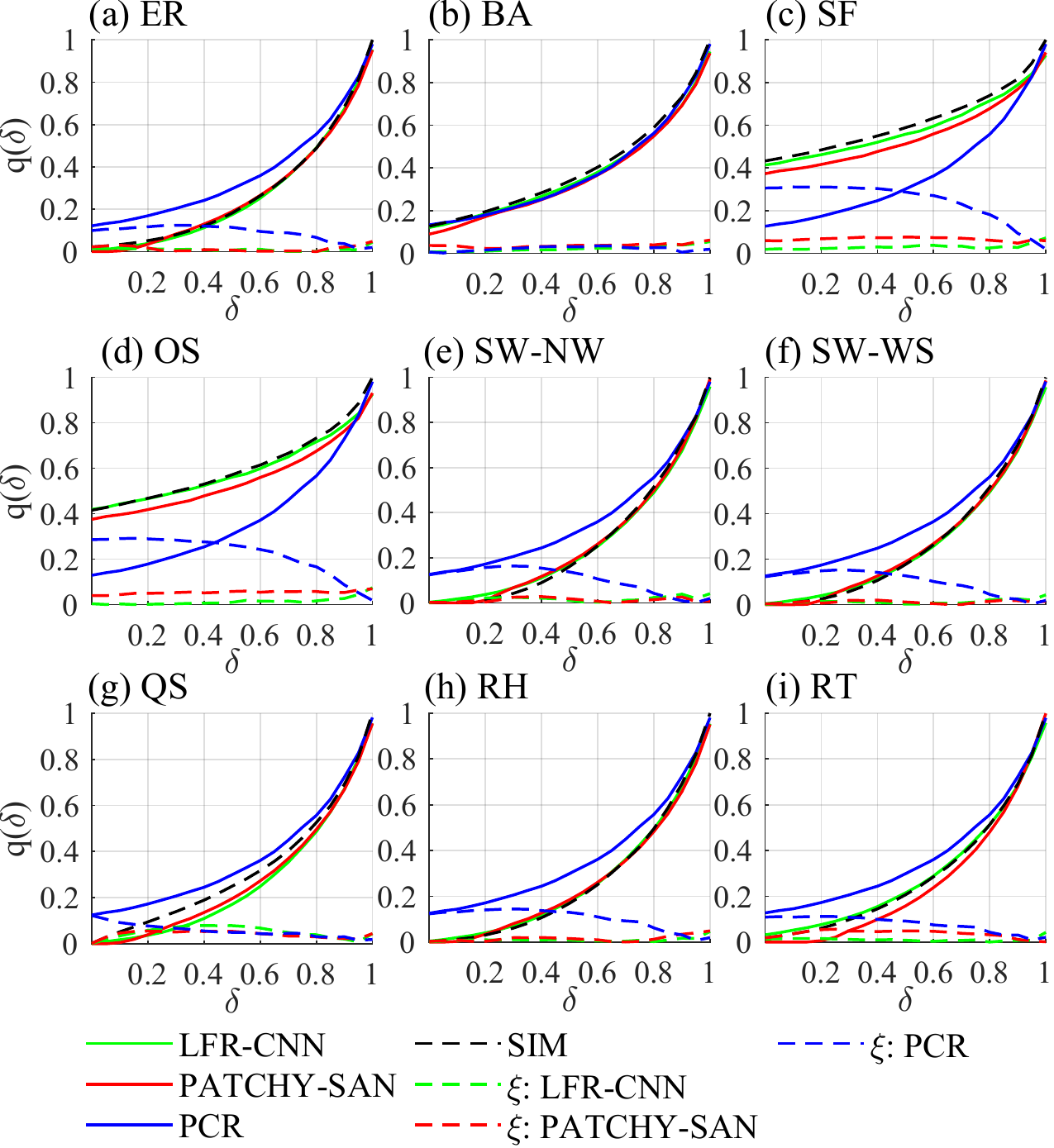}
	\caption{[color online] Comparison of prediction results using LFR-CNN, PCR, and PATCHY-SAN, for controllability robustness of directed networks ($N\in[350,650]$) under RA.}\label{fig:yc_d_rnd}
\end{figure}
\begin{figure}[htbp]
	\centering
	\includegraphics[width=\linewidth]{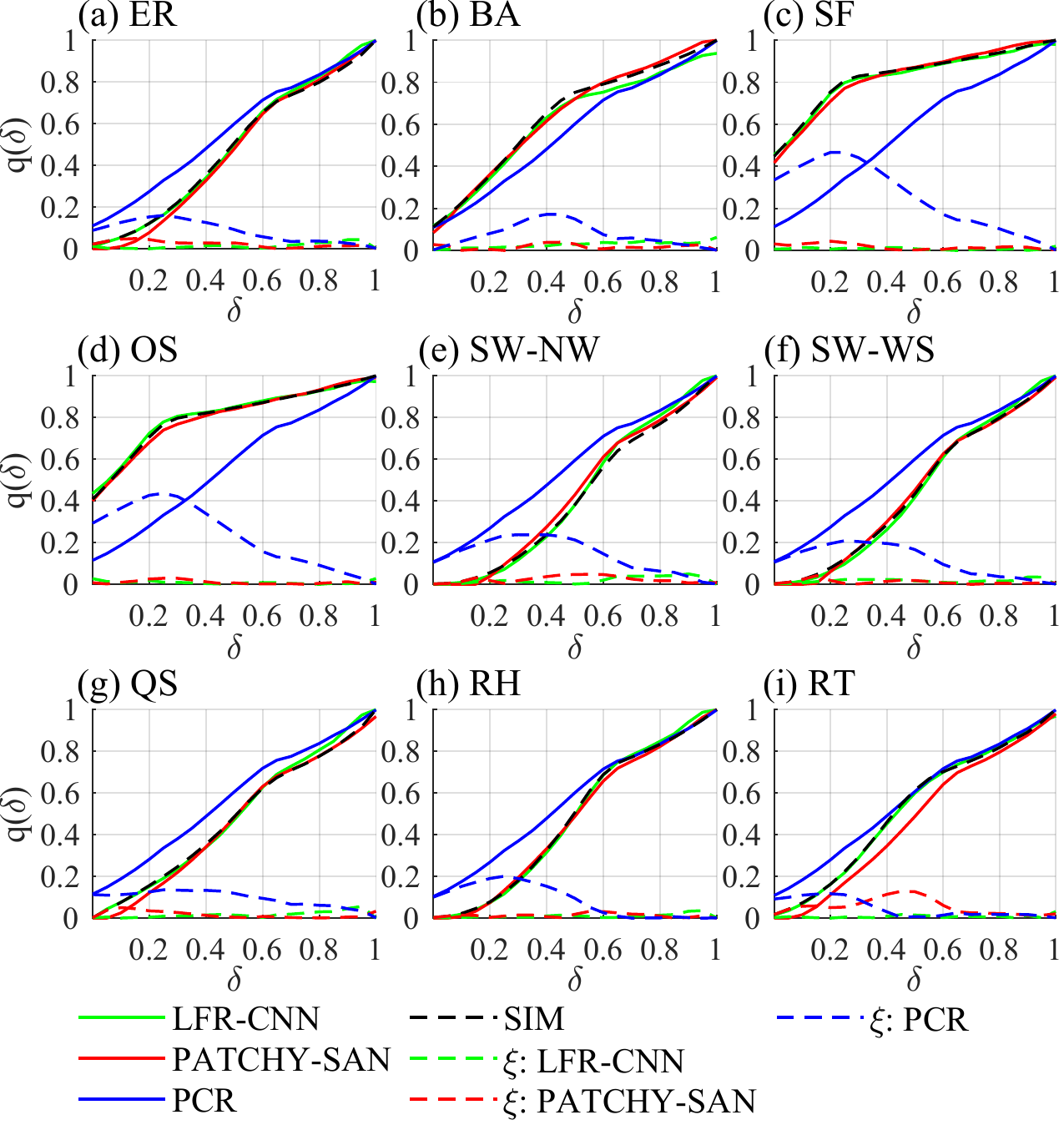}\\
	\caption{[color online] Comparison of prediction results using LFR-CNN, PCR, and PATCHY-SAN, for controllability robustness of directed networks ($N\in[350,650]$) under TB.}\label{fig:yc_d_bet}
\end{figure}

Controllability robustness of directed networks under RA and TB is predicted using LFR-CNN, PCR, and PATCHY-SAN. The simulation results in terms of controllability curves are shown in Figs. \ref{fig:yc_d_rnd} and \ref{fig:yc_d_bet}, respectively. A network controllability curve is denoted by $q(\delta)$, where $\delta$ represents the proportion of removed nodes. For each predictor, its predicted controllability curve and prediction error curve are plotted in the same color; `SIM' denotes the controllability curve obtained by attack simulations. Each curve is averaged from 100 testing samples.

As shown in Figs. \ref{fig:yc_d_rnd} and \ref{fig:yc_d_bet}, PCR performs badly in prediction. This is due to the following two reasons: 1) both the training and testing data have a wide network size variation with $N\in[350,650]$ and $\langle k\rangle\in[1.5,6]$; and 2) there 9 synthetic network types trained and tested. As a result, PCR predicts the controllability curves almost in the same pattern for all different networks with different sizes and average degrees. In contrast, LFR-CNN and PATCHY-SAN, both contain an LFR module, are able to predict different controllability curves for different scenarios. In Figs. \ref{fig:yc_d_rnd} (c), (d), (i), and Fig. \ref{fig:yc_d_bet} (i), it is visible that the green curves (LFR-CNN predictions) are closer to the black dotted curves (true simulation results) than the red curves (PATCHY-SAN predictions), meaning that LFR-CNN performs clearly better than PATCHY-SAN in prediction.

Table \ref{tab:main_res} summarizes the overall prediction errors of the three predictors in different experiments, with Kruskal-Wallis H-test \cite{Kruskal1952JASA} results. The overall errors of the results in Fig. \ref{fig:yc_d_rnd} are shown in Table \ref{tab:main_res} (I), which shows that 1) LFR-CNN performs significantly better than PCR for all networks; 2) LFR-CNN performs significantly better than PATCHY-SAN for ER, SF, OS, and RT, but significantly worse than PATCHY-SAN for SW-WS, QS, and RH. The overall errors of the results in Fig. \ref{fig:yc_d_bet} are shown in Table \ref{tab:main_res} (II), which shows that 1) LFR-CNN performs significantly better than PCR for all networks; 2) LFR-CNN performs significantly better than PATCHY-SAN for ER, SW-NW, SW-WS, RH, and RT, but significantly worse than PATCHY-SAN for BA. All in all, LFR-CNN outperforms PCR for all networks; LFR-CNN outperforms PATCHY-SAN in 9 comparisons, but is worse in 4 comparisons, while in the other 5 comparisons, two predictors have no significant differences.

\begin{table*}[htbp]
	\centering
	\caption{Comparison of average prediction errors among LFR-CNN, PCR and PATCHY-SAN, where $N\in[350,650]$. The signs in parentheses denote the Kruskal-Wallis H-test \cite{Kruskal1952JASA} results of LFR-CNN vs PCR and LFR-CNN vs PATCHY-SAN, respectively. A `$+$' sign denotes that LFR-CNN significantly outperforms the other method by obtaining lower errors; a `$\approx$' sign denotes no significant difference between two methods; and a `$-$' sign denotes that LFR-CNN performs significantly worse than the other methods with greater errors. }
	\begin{tabular}{|lc|c|c|c|c|c|c|c|c|c|} \hline
		\multicolumn{2}{|c|}{Average Prediction Error $\bar{\xi}$} & ER & BA & SF & OS & SW-NW & SW-WS & QS & RH & RT \\ \hline
		\multicolumn{1}{|l|}{\multirow{3}{*}{\begin{tabular}[c]{@{}l@{}}(I) Controllability\\Robustness of Directed\\Networks under RA \end{tabular}}} & LFR-CNN & \begin{tabular}[c]{@{}c@{}}0.0450\\ ($+$,$+$)\end{tabular} & \begin{tabular}[c]{@{}c@{}}0.0395\\ ($+$,$\approx$)\end{tabular} & \begin{tabular}[c]{@{}c@{}}0.0601\\ ($+$,$+$)\end{tabular} & \begin{tabular}[c]{@{}c@{}}0.0567\\($+$,$+$)\end{tabular}& \begin{tabular}[c]{@{}c@{}}0.0480\\($+$,$\approx$)\end{tabular} & \begin{tabular}[c]{@{}c@{}}0.0361\\($+$,$-$)\end{tabular} & \begin{tabular}[c]{@{}c@{}}0.0375\\($+$,$-$)\end{tabular}& \begin{tabular}[c]{@{}c@{}}0.0440\\($+$,$-$)\end{tabular} & \begin{tabular}[c]{@{}c@{}}0.0474\\($+$,$+$)\end{tabular}\\ \cline{2-11}
		\multicolumn{1}{|l|}{} & PCR & 0.1280 & 0.1509& 0.2689 & 0.2541 & 0.1139 & 0.1358 & 0.1301 & 0.1331 & 0.1360\\ \cline{2-11}
		\multicolumn{1}{|l|}{} & PATCHY-SAN & 0.0313 & 0.0458& 0.0732& 0.0601& 0.0450 & 0.0253 & 0.0272& 0.0304 & 0.0541\\ \hline
		\multicolumn{1}{|l|}{\multirow{3}{*}{\begin{tabular}[c]{@{}l@{}}(II) Controllability\\Robustness of Directed\\Networks under TB\end{tabular}}} & LFR-CNN & \begin{tabular}[c]{@{}c@{}}0.02544\\($+$,$+$)\end{tabular} & \begin{tabular}[c]{@{}c@{}}0.05219\\($+$,$-$)\end{tabular} & \begin{tabular}[c]{@{}c@{}}0.04376\\($+$,$\approx$)\end{tabular} & \begin{tabular}[c]{@{}c@{}}0.04650\\($+$,$\approx$)\end{tabular} & \begin{tabular}[c]{@{}c@{}}0.02355\\($+$,$+$)\end{tabular} & \begin{tabular}[c]{@{}c@{}}0.02445\\($+$,$+$)\end{tabular} & \begin{tabular}[c]{@{}c@{}}0.02210\\($+$,$\approx$)\end{tabular} & \begin{tabular}[c]{@{}c@{}}0.02134\\($+$,$+$)\end{tabular} & \begin{tabular}[c]{@{}c@{}}0.03641\\($+$,$+$)\end{tabular} \\ \cline{2-11}
		\multicolumn{1}{|l|}{} & PCR & 0.1369 & 0.1625 & 0.2704& 0.2570& 0.1374 & 0.1548 & 0.1384& 0.1302 & 0.1300 \\ \cline{2-11}
		\multicolumn{1}{|l|}{} & PATCHY-SAN & 0.0354 & 0.0351& 0.0391& 0.0388& 0.0273 & 0.0333 & 0.0238& 0.0258 & 0.0614\\ \hline
		\multicolumn{1}{|l|}{\multirow{3}{*}{\begin{tabular}[c]{@{}l@{}}(III) Connectivity\\Robustness of Undirected\\Networks under RA\end{tabular}}} & LFR-CNN & \begin{tabular}[c]{@{}c@{}}0.0362\\($+$,$+$)\end{tabular} & \begin{tabular}[c]{@{}c@{}}0.0665\\($\approx$,$\approx$)\end{tabular} & \begin{tabular}[c]{@{}c@{}}0.0868\\($+$,$\approx$)\end{tabular} & \begin{tabular}[c]{@{}c@{}}0.0908\\($+$,$\approx$)\end{tabular} & \begin{tabular}[c]{@{}c@{}}0.0338\\($+$,$+$)\end{tabular}& \begin{tabular}[c]{@{}c@{}}0.0365\\($+$,$+$)\end{tabular} & \begin{tabular}[c]{@{}c@{}}0.0350\\($+$,$+$)\end{tabular}& \begin{tabular}[c]{@{}c@{}}0.0406\\($+$,$+$)\end{tabular} & \begin{tabular}[c]{@{}c@{}}0.0767\\($\approx$,$\approx$)\end{tabular} \\ \cline{2-11}
		\multicolumn{1}{|l|}{} & PCR & 0.0695 & 0.0767& 0.1167& 0.1219& 0.0663 & 0.0863 & 0.0825& 0.0728 & 0.0779\\ \cline{2-11}
		\multicolumn{1}{|l|}{} & PATCHY-SAN & 0.0639 & 0.0692& 0.0835& 0.0803& 0.0703 & 0.0670 & 0.0663& 0.0590 & 0.0635\\ \hline
		\multicolumn{1}{|l|}{\multirow{3}{*}{\begin{tabular}[c]{@{}l@{}}(IV) Connectivity\\Robustness of Undirected\\Networks under TD\end{tabular}}} & LFR-CNN & \begin{tabular}[c]{@{}c@{}}0.0302\\($+$,$+$)\end{tabular} & \begin{tabular}[c]{@{}c@{}}{0.0334}\\($+$,$\approx$)\end{tabular}& \begin{tabular}[c]{@{}c@{}}0.0215\\($+$,$\approx$)\end{tabular} & \begin{tabular}[c]{@{}c@{}}0.0262\\($+$,$\approx$)\end{tabular} & \begin{tabular}[c]{@{}c@{}}0.0279\\($+$,$+$)\end{tabular}& \begin{tabular}[c]{@{}c@{}}0.0265\\($+$,$+$)\end{tabular}& \begin{tabular}[c]{@{}c@{}}0.0254\\($+$,$+$)\end{tabular}& \begin{tabular}[c]{@{}c@{}}0.0345\\($+$,$+$)\end{tabular}& \begin{tabular}[c]{@{}c@{}}0.0563\\($+$,$\approx$)\end{tabular}\\ \cline{2-11}
		\multicolumn{1}{|l|}{} & PCR & 0.1423 & 0.1680& 0.2724& 0.2792& 0.1644 & 0.1520 & 0.1402& 0.1351 & 0.1386\\ \cline{2-11}
		\multicolumn{1}{|l|}{} & PATCHY-SAN & 0.0404 & 0.0420& 0.0230& 0.0282& 0.0501 & 0.0446 & 0.0439& 0.0408 & 0.0460\\ \hline
	\end{tabular}\label{tab:main_res}
\end{table*}

\subsection{Predicting Controllability Robustness for Real-world Networks} \label{sub:exp_real}

\begin{figure}[htbp]
	\centering
	\includegraphics[width=\linewidth]{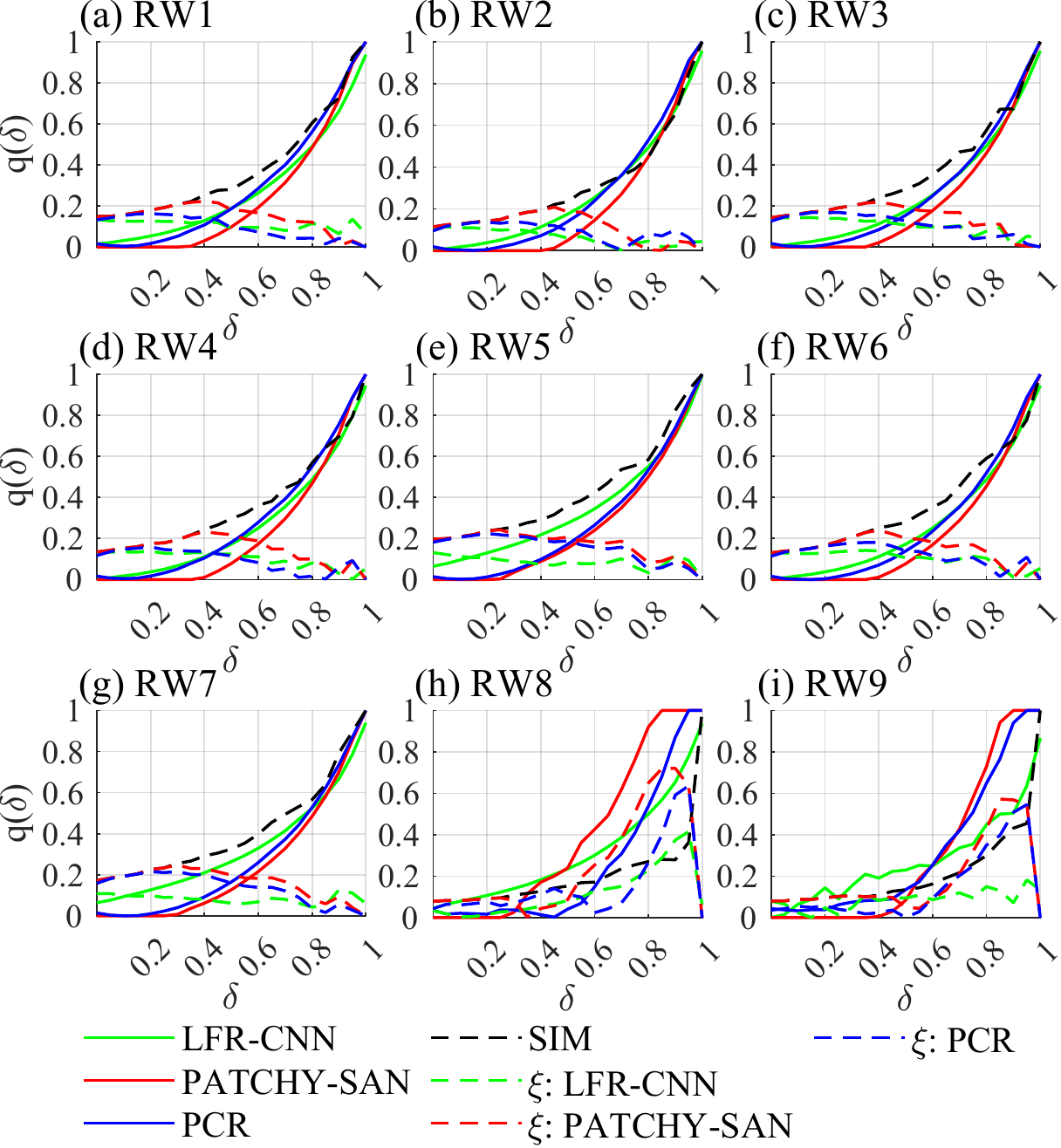}
	\caption{[color online] Comparison of prediction results using LFR-CNN, PCR, and PATCHY-SAN, for controllability robustness of REDDIT-MULTI \cite{Yanardag2015KDD} real-world networks ($N\in[419,570]$) under RA.}\label{fig:yc_rnd_real}
\end{figure}
\begin{table*}[htbp]
	\centering
	\caption{Basic Information of REDDIT-MULTI real-world networks. Comparison of average prediction errors among LFR-CNN, PCR and PATCHY-SAN, where $N\in[419,570]$. Numbers in parentheses denote the ranks of predictors in ascending order of prediction errors. }
	\begin{tabular}{|cccccccccc|}
		\hline
		\multicolumn{1}{|c|}{}& \multicolumn{1}{c|}{RW1} & \multicolumn{1}{c|}{RW2}& \multicolumn{1}{c|}{RW3}& \multicolumn{1}{c|}{RW4}& \multicolumn{1}{c|}{RW5}& \multicolumn{1}{c|}{RW6}& \multicolumn{1}{c|}{RW7}& \multicolumn{1}{c|}{RW8}& RW9\\ \hline
		\multicolumn{1}{|c|}{REDDIT-MULTI \cite{Yanardag2015KDD}} & \multicolumn{1}{c|}{12K-16} & \multicolumn{1}{c|}{12K-40} & \multicolumn{1}{c|}{12K-41} & \multicolumn{1}{c|}{12K-49} & \multicolumn{1}{c|}{12K-81} & \multicolumn{1}{c|}{12K-124}& \multicolumn{1}{c|}{12K-129}& \multicolumn{1}{c|}{5K-1} & 5K-2\\ \hline
		\multicolumn{1}{|c|}{$N$} & \multicolumn{1}{c|}{499}& \multicolumn{1}{c|}{510}& \multicolumn{1}{c|}{538}& \multicolumn{1}{c|}{551}& \multicolumn{1}{c|}{499}& \multicolumn{1}{c|}{522}& \multicolumn{1}{c|}{570}& \multicolumn{1}{c|}{419}& 428\\ \hline
		\multicolumn{1}{|c|}{$\langle k\rangle$} & \multicolumn{1}{c|}{6.31} & \multicolumn{1}{c|}{8.93} & \multicolumn{1}{c|}{6.84} & \multicolumn{1}{c|}{7.15} & \multicolumn{1}{c|}{4.95} & \multicolumn{1}{c|}{7.56} & \multicolumn{1}{c|}{5.75} & \multicolumn{1}{c|}{47.07} & 35.01 \\ \hline \hline
		\multicolumn{1}{|c|}{LFR-CNN}& \multicolumn{1}{c|}{0.1082 (2)} & \multicolumn{1}{c|}{0.0667 (1)} & \multicolumn{1}{c|}{0.1035 (1)} & \multicolumn{1}{c|}{0.1014 (2)} & \multicolumn{1}{c|}{0.0856 (1)} & \multicolumn{1}{c|}{0.1041 (1)} & \multicolumn{1}{c|}{0.0824 (1)} & \multicolumn{1}{c|}{0.1168 (1)} & 0.0875 (1) \\ \hline
		\multicolumn{1}{|c|}{PCR}& \multicolumn{1}{c|}{0.0969 (1)} & \multicolumn{1}{c|}{0.0938 (2)} & \multicolumn{1}{c|}{0.1104 (2)} & \multicolumn{1}{c|}{0.0949 (1)} & \multicolumn{1}{c|}{0.1532 (2)} & \multicolumn{1}{c|}{0.1224 (2)} & \multicolumn{1}{c|}{0.1378 (2)} & \multicolumn{1}{c|}{0.1866 (2)} & 0.1718 (2) \\ \hline
		\multicolumn{1}{|c|}{PATCHY-SAN}& \multicolumn{1}{c|}{0.1503 (3)} & \multicolumn{1}{c|}{0.1211 (3)} & \multicolumn{1}{c|}{0.1497 (3)} & \multicolumn{1}{c|}{0.1531 (3)} & \multicolumn{1}{c|}{0.1733 (3)} & \multicolumn{1}{c|}{0.1563 (3)} & \multicolumn{1}{c|}{0.1679 (3)} & \multicolumn{1}{c|}{0.3611 (3)} & 0.2636 (3) \\ \hline
	\end{tabular}\label{tab:rwn}
\end{table*}

A total of 9 real-world network instances are randomly selected from the Reddit multiset data \cite{Yanardag2015KDD}. Three predictors are used to predict the controllability robustness of there real-world networks under RA. The basic information of these networks and the prediction errors obtained by the three predictors are summarized in Table \ref{tab:rwn}. Ranks of predictors in ascending order are attached in parentheses following the prediction errors, where the average ranks of LFR-CNN, PCR and PATCHY-SAN are 1.22, 1.78, and 3, respectively. This suggests that LFR-CNN and PCR have better generalizability than PATCHY-SAN for unknown real-world networks, although the overall prediction errors for all three predictors are relatively greater than that for synthetic networks. The predicted controllability curves are shown in Fig. \ref{fig:yc_rnd_real}, which demonstrate that LFR-CNN predicts the controllability curves closer to the simulation results than the other two predictors.

\subsection{Predicting Connectivity Robustness for Undirected Networks} \label{sub:exp_und}

\begin{figure}[htbp]
	\centering
	\includegraphics[width=\linewidth]{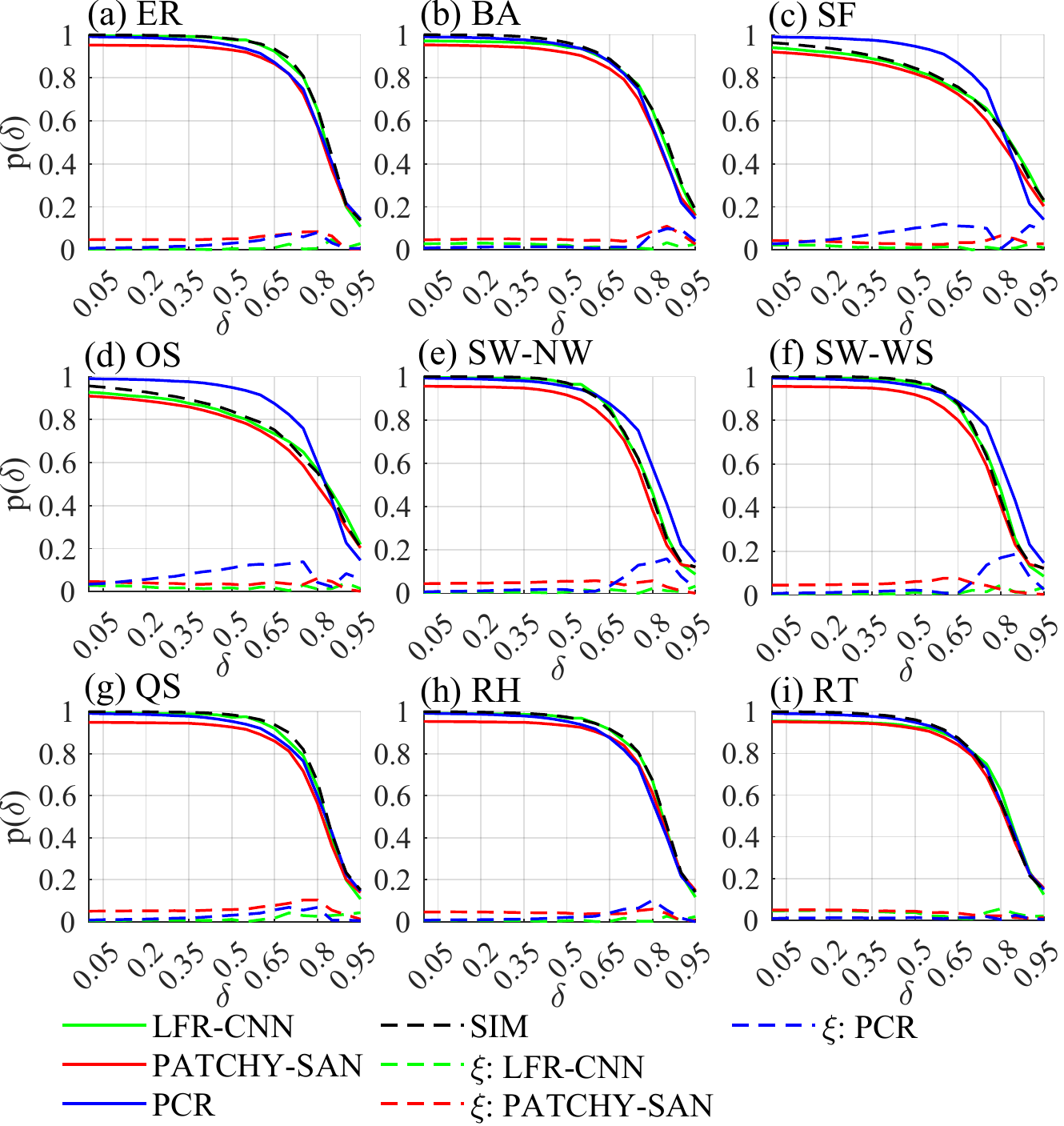}
	\caption{[color online] Comparison of prediction results using LFR-CNN, PCR, and PATCHY-SAN, for connectivity robustness of undirected networks ($N\in[350,650]$) under RA.}\label{fig:lc_ud_rnd}
\end{figure}	
\begin{figure}[htbp]
	\centering
	\includegraphics[width=\linewidth]{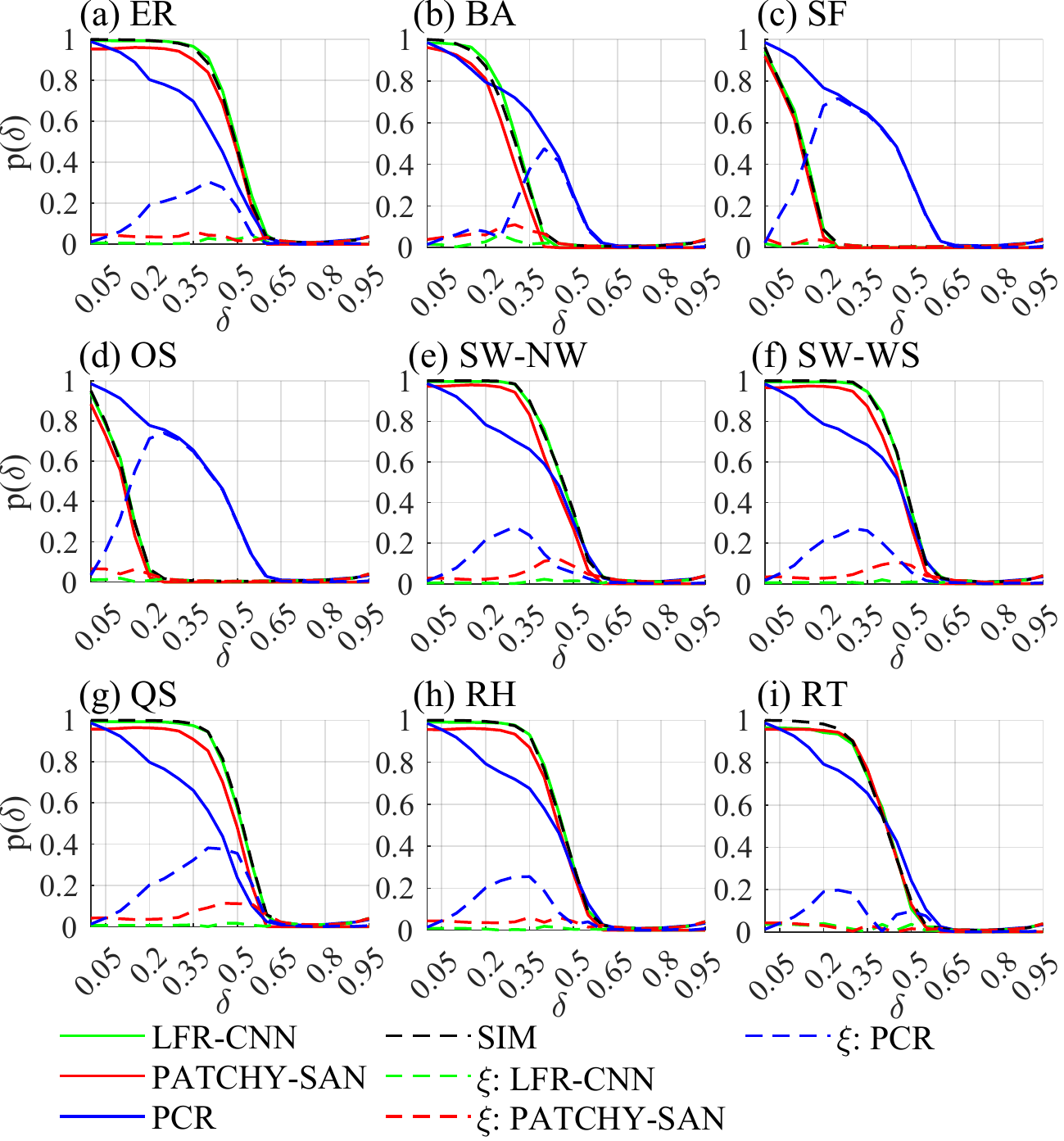}
	\caption{[color online] Comparison of prediction results using LFR-CNN, PCR, and PATCHY-SAN, for connectivity robustness of undirected networks ($N\in[350,650]$) under TD.}\label{fig:lc_ud_deg}
\end{figure}	

CNN-based approaches are capable of dealing with \textit{all} types of complex networks, including weighted and unweighted, directed and undirected, real-world and synthetic networks \cite{Lou2021TNNLS}. Here, for brevity, a comparison of connectivity robustness predictions is performed only on undirected networks. The predicted connectivity curves under RA are shown in Fig. \ref{fig:lc_ud_rnd}, for which the overall prediction errors are summarized in Table \ref{tab:main_res} (III). Figure \ref{fig:lc_ud_rnd} shows that all the three predictors perform well (or fairly good) on predicting the connectivity curves, which are denoted by $p(\delta)$. Table \ref{tab:main_res} (III) shows that the prediction errors are mostly in a magnitude of $10^{-2}$. The predicted curves under TD are shown in Fig. \ref{fig:lc_ud_deg}, for which the overall errors are summarized in Table \ref{tab:main_res} (IV). It is clear that PCR performs imprecisely well.

The data summarized in Tables \ref{tab:main_res} (III) and (IV) suggest that LFR-CNN outperforms PCR and PATCHY-SAN in predicting 16 out of 18 and 10 out of 18 comparisons, respectively, while for the rest networks, LFR-CNN performs statistically equivalently well as PCR and PATCHY-SAN.

In a nutshell, LFR-CNN outperforms PCR in 34/36 cases, and outperforms PACTHY-SAN in 19/36 cases; PACTHY-SAN outperforms LFR-CNN in 4/36 cases, while PCR does not outperform LFR-CNN in any case; for the rest cases, no significant differences are detected.

\subsection{Node Attributes as Receptive Fields} \label{sub:exp_ft}

\begin{table*}[htbp]
	\centering \caption{Comparison of average prediction errors obtained using different attribute combinations in LFR-CNN. Three node attributes, including degree ({\MakeLowercase{\textit{deg}}}), clustering coefficient ({\MakeLowercase{\textit{cc}}}), and betweenness ({\MakeLowercase{\textit{bet}}}), compose three pairwise combinations.}
	\begin{tabular}{|lc|c|c|c|c|c|c|c|c|c|} \hline
		\multicolumn{2}{|l|}{} & ER & BA & SF & OS & SW-NW & SW-WS & QS & RH & RT \\ \hline
		\multicolumn{1}{|l|}{\multirow{3}{*}{\begin{tabular}[c]{@{}l@{}}(I) Controllability\\Robustness of Directed\\ Networks under RA\end{tabular}}} & \textit{deg} \& \textit{cc} & \begin{tabular}[c]{@{}c@{}}0.0432\\ ($\approx$,$+$)\end{tabular} & \begin{tabular}[c]{@{}c@{}}0.0357\\ ($+$,$+$)\end{tabular} & \begin{tabular}[c]{@{}c@{}}0.0436\\ ($\approx$,$+$)\end{tabular} & \begin{tabular}[c]{@{}c@{}}0.0372\\ ($+$,$+$)\end{tabular} & \begin{tabular}[c]{@{}c@{}}0.0581\\ ($+$,$+$)\end{tabular} & \begin{tabular}[c]{@{}c@{}}0.0322\\ ($\approx$,$+$)\end{tabular} & \begin{tabular}[c]{@{}c@{}}0.0351\\ ($\approx$,$+$)\end{tabular} & \begin{tabular}[c]{@{}c@{}}0.0399\\ ($\approx$,$+$)\end{tabular} & \begin{tabular}[c]{@{}c@{}}0.0421\\ ($+$,$+$)\end{tabular} \\ \cline{2-11}
		\multicolumn{1}{|l|}{} & \textit{deg} \& \textit{bet} & 0.0384 & 0.0562 & 0.0472 & 0.0556 & 0.0439 & 0.0321 & 0.0337 & 0.0394 & 0.0515 \\ \cline{2-11}
		\multicolumn{1}{|l|}{} & \textit{bet} \& \textit{cc} & 0.0589 & 0.0865 & 0.1203 & 0.1179 & 0.0640 & 0.0543 & 0.0566 & 0.0571 & 0.0681 \\ \hline
		\multicolumn{1}{|l|}{\multirow{3}{*}{\begin{tabular}[c]{@{}l@{}}(II) Connectivity\\ Robustness of Undirected\\ Networks under RA\end{tabular}}} & \textit{deg} \& \textit{cc} & \begin{tabular}[c]{@{}c@{}}0.0293\\ ($+$,$+$)\end{tabular} & \begin{tabular}[c]{@{}c@{}}0.0490\\ ($\approx$,$+$)\end{tabular} & \begin{tabular}[c]{@{}c@{}}0.0791\\ ($\approx$,$+$)\end{tabular} & \begin{tabular}[c]{@{}c@{}}0.0769\\ ($+$,$+$)\end{tabular} & \begin{tabular}[c]{@{}c@{}}0.0287\\ ($+$,$+$)\end{tabular} & \begin{tabular}[c]{@{}c@{}}0.0288\\ ($+$,$+$)\end{tabular} & \begin{tabular}[c]{@{}c@{}}0.0287\\ ($+$,$+$)\end{tabular} & \begin{tabular}[c]{@{}c@{}}0.0340\\ ($+$,$+$)\end{tabular} & \begin{tabular}[c]{@{}c@{}}0.0461\\ ($\approx$,$+$)\end{tabular} \\ \cline{2-11}
		\multicolumn{1}{|l|}{} & \textit{deg} \& \textit{bet} & 0.0503 & 0.0494 & 0.0921 & 0.0937 & 0.0635 & 0.0562 & 0.0527 & 0.0508 & 0.0568 \\ \cline{2-11}
		\multicolumn{1}{|l|}{} & \textit{bet} \& \textit{cc} & 0.1291 & 0.1434 & 0.1628 & 0.1632 & 0.1331 & 0.1339 & 0.1298 & 0.1325 & 0.1454 \\ \hline
	\end{tabular}\label{tab:atb}
\end{table*}

In the normalization step of LFR, the attributes of the selected neighborhood nodes are embedded in a receptive field. Here, different combinations of node attributes including degree, clustering coefficient, and betweenness are compared. Table \ref{tab:atb} shows the prediction errors for (I) controllability robustness and (II) connectivity robustness, among the three combinations. It is clear that the default setting using degree and clustering coefficient (\textit{deg} \& \textit{cc}) outperforms the other two combinations.

\subsection{Scalability of Network Size} \label{sub:exp_size}

\begin{figure}[htbp]
	\centering
	\includegraphics[width=\linewidth]{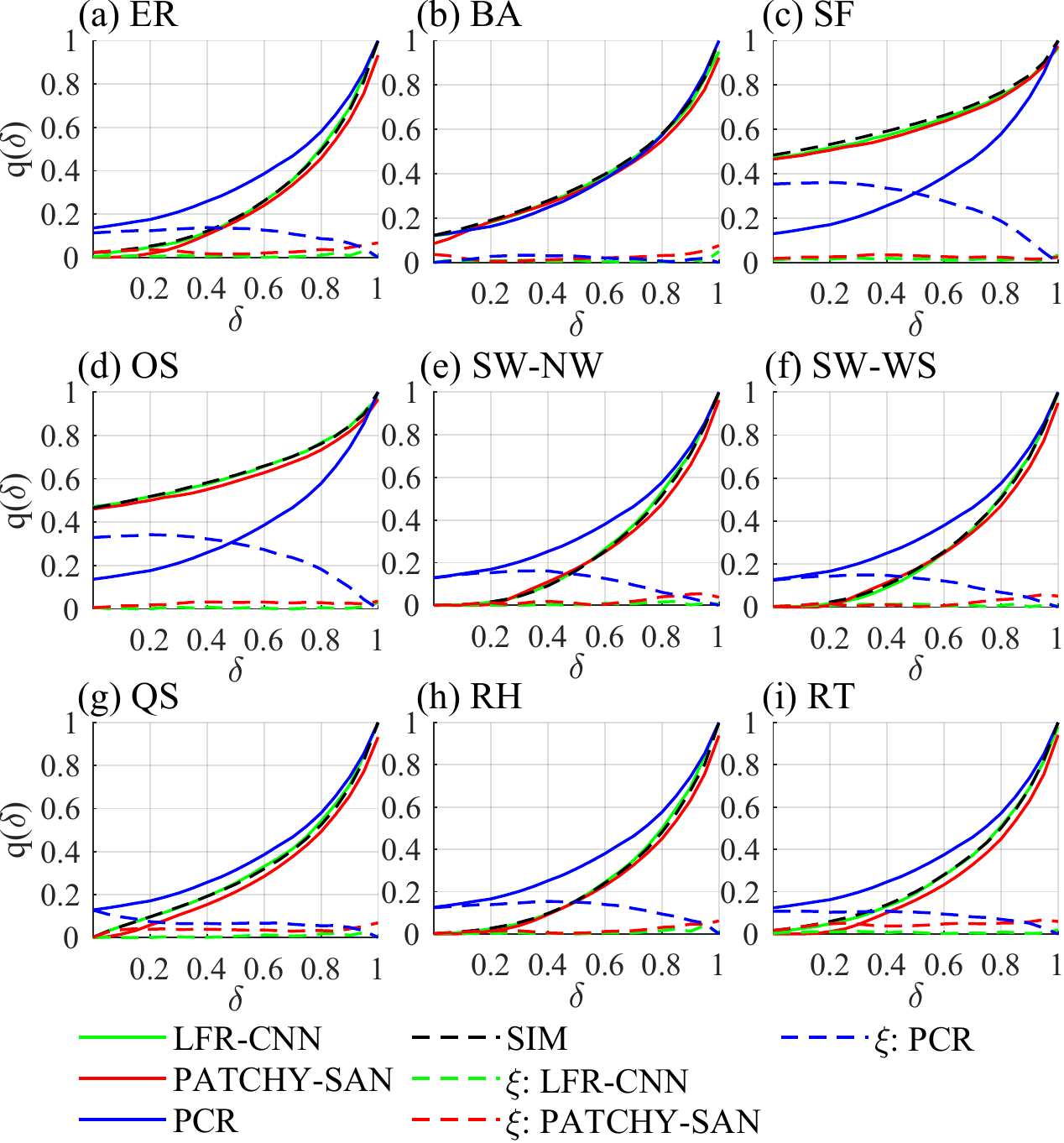}
	\caption{[color online] Comparison of prediction results using LFR-CNN, PCR, and PATCHY-SAN, for controllability robustness of directed networks ($N\in[700,1300]$) under RA.}\label{fig:yc_d_rnd_1k}
\end{figure}
\begin{figure}[htbp]
	\centering
	\includegraphics[width=\linewidth]{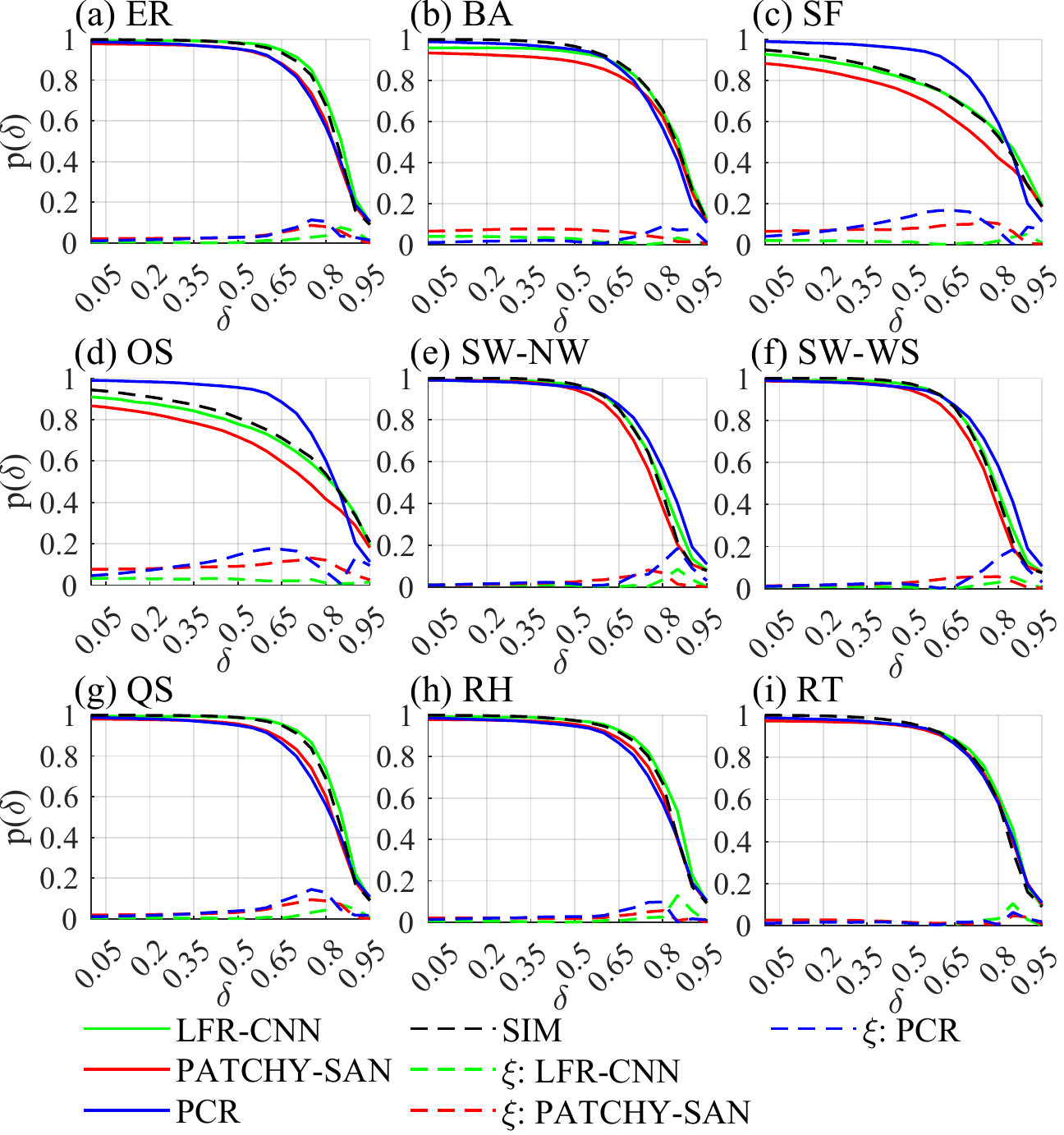}
	\caption{[color online] Comparison of prediction results using LFR-CNN, PCR, and PATCHY-SAN, for connectivity robustness of undirected networks ($N\in[700,1300]$) under RA.}\label{fig:lc_ud_rnd_1k}
\end{figure}
\begin{table*}[htbp]
	\centering
	\caption{Comparison of average prediction errors among LFR-CNN, PCR and PATCHY-SAN, where $N\in[700,1300]$. The signs in parentheses denote the Kruskal-Wallis H-test \cite{Kruskal1952JASA} results of LFR-CNN vs PCR and LFR-CNN vs PATCHY-SAN, respectively. A `$+$' sign denotes that LFR-CNN significantly outperforms the other method by obtaining lower errors; a `$\approx$' sign denotes no significant difference between two methods; and a `$-$' sign denotes that LFR-CNN performs significantly worse than the other methods with greater errors. }
	\begin{tabular}{|cc|c|c|c|c|c|c|c|c|c|}
		\hline \multicolumn{2}{|c|}{Average Prediction Error $\bar{\xi}$} & ER & BA & SF & OS & SW-NW & SW-WS & QS & RH & RT\\ \hline
		\multicolumn{1}{|c|}{\multirow{3}{*}{\begin{tabular}[c]{@{}l@{}}(I) Controllability\\Robustness of Directed\\Networks under RA\end{tabular}}} & \begin{tabular}[c]{@{}c@{}}LFR-CNN\end{tabular} & \begin{tabular}[c]{@{}c@{}}0.0191\\($+$,$+$)\end{tabular} & \begin{tabular}[c]{@{}c@{}}0.0406\\($+$,$\approx$)\end{tabular} & \begin{tabular}[c]{@{}c@{}}0.0356\\($+$,$\approx$)\end{tabular} & \begin{tabular}[c]{@{}c@{}}0.0341\\($+$,$\approx$)\end{tabular} & \begin{tabular}[c]{@{}c@{}}0.0151\\($+$,$+$)\end{tabular} & \begin{tabular}[c]{@{}c@{}}0.0171\\($+$,$+$)\end{tabular} & \begin{tabular}[c]{@{}c@{}}0.0162\\($+$,$+$)\end{tabular} & \begin{tabular}[c]{@{}c@{}}0.0177\\($+$,$+$)\end{tabular} & \begin{tabular}[c]{@{}c@{}}0.0316\\($+$,$+$)\end{tabular} \\ \cline{2-11}
		\multicolumn{1}{|c|}{}& PCR & 0.1433& 0.1408 & 0.2820 & 0.2706 & 0.1349& 0.1282 & 0.1242& 0.1395& 0.1284 \\ \cline{2-11}
		\multicolumn{1}{|c|}{} & PATCHY-SAN & 0.0374& 0.0387 & 0.0420 & 0.0448 & 0.0259& 0.0240 & 0.0375& 0.0268 & 0.0499 \\ \hline
		\multicolumn{1}{|c|}{\multirow{3}{*}{\begin{tabular}[c]{@{}l@{}}(II) Connectivity\\Robustness of Undirected\\Networks under RA\end{tabular}}} & \begin{tabular}[c]{@{}c@{}}LFR-CNN\end{tabular} & \begin{tabular}[c]{@{}c@{}}0.0266\\($+$,$+$)\end{tabular} & \begin{tabular}[c]{@{}c@{}}0.0594\\($\approx$,$+$)\end{tabular} & \begin{tabular}[c]{@{}c@{}}0.0705\\($+$,$+$)\end{tabular} & \begin{tabular}[c]{@{}c@{}}0.0790\\($+$,$+$)\end{tabular} & \begin{tabular}[c]{@{}c@{}}0.0239\\($+$,$+$)\end{tabular} & \begin{tabular}[c]{@{}c@{}}0.0297\\($+$,$\approx$)\end{tabular} & \begin{tabular}[c]{@{}c@{}}0.0271\\($+$,$+$)\end{tabular} & \begin{tabular}[c]{@{}c@{}}0.0293\\($+$,$+$)\end{tabular} & \begin{tabular}[c]{@{}c@{}}0.0424\\($+$,$\approx$)\end{tabular} \\ \cline{2-11}
		\multicolumn{1}{|c|}{}& PCR & 0.0654 & 0.0744 & 0.1321 & 0.1348 & 0.0784 & 0.0861 & 0.0833 & 0.0741& 0.0809 \\ \cline{2-11}
		\multicolumn{1}{|c|}{}& PATCHY-SAN & 0.0440 & 0.0757 & 0.0971 & 0.1070 & 0.0479 & 0.0444 & 0.0427 & 0.0357 & 0.0398 \\ \hline
	\end{tabular}\label{tab:res_1k}
\end{table*}

To further verify the scalability, the proposed LFR-CNN is compared with PCR and PATCHY-SAN on predicting a set of networks of sizes $N\in[700,1300]$. Here, a 7-FM PCR is employed and $W=1000$ is set for LFR-CNN and PATCHY-SAN.

The predicted controllability and connectivity curves under RA are shown in Figs. \ref{fig:yc_d_rnd_1k} and \ref{fig:lc_ud_rnd_1k}, respectively. It is visible that LFR-CNN and PATCHY-SAN perform better than PCR in controllability robustness prediction. As for connectivity robustness, LFR-CNN performs visibly better than PATCHY-SAN and PCR in Figs. \ref{fig:lc_ud_rnd_1k} (c) and (d).

The overall prediction errors are shown in Table \ref{tab:res_1k}. LFR-CNN outperforms PCR for 17 out of 18 cases, and outperforms PATCHY-SAN for 13 out of 18 cases; while for the rest comparisons, LFR-CNN performs statistically equivalently to PCR or PATCHY-SAN in prediction.

\subsection{Network Size Variation}
\label{sub:exp_pcr}

\begin{table}[htbp]
	\centering
	\caption{Comparison of average prediction errors among LFR-CNN, PCR and PATCHY-SAN, where $N=800$.}
	\begin{tabular}{|c|c|c|c|c|}
		\hline
		\begin{tabular}[c]{@{}c@{}}Average Prediction\\Error $\bar{\xi}$\end{tabular} & ER & SF & QS& SW-NW\\ \hline
		LFR-CNN & \begin{tabular}[c]{@{}c@{}}0.0189\\ ($\approx$,$+$)\end{tabular} & \begin{tabular}[c]{@{}c@{}}0.0750\\ ($-$,$+$)\end{tabular} & \begin{tabular}[c]{@{}c@{}}0.0162\\ ($\approx$,$+$)\end{tabular} & \begin{tabular}[c]{@{}c@{}}0.0157\\ ($\approx$,$+$)\end{tabular} \\ \hline
		PCR & 0.0166& 0.0194 & 0.0145& 0.0141 \\ \hline
		PATCHY-SAN & 0.0253& 0.1074 & 0.0208& 0.0263\\ \hline
	\end{tabular}\label{tab:8}
\end{table}

The core prediction component in LFR-CNN, PCR, and PATCHY-SAN is a 3-FM CNN, a 7-FM CNN, and a 1D-CNN, respectively. These CNN-based core components perform the regression task and predict the robustness performance for an input network. In PCR, the input data to CNN are adjacency matrices, while for LFR-CNN and PATCHY-SAN, the LFR module will convert the raw adjacency matrices to lower-dimensional representations before inputting them to the respective CNNs. Specifically, suppose that $H$ is the input size of the prediction component of LFR-CNN, PCR, or PATCHY-SAN, and given an input adjacency matrix of size $J\times J$ ($J\neq H$). Upsampling or downsampling is necessary to resize the input for PCR, where the original adjacency information may be significantly modified. In contrast, for LFR-CNN and PATCHY-SAN, the $J\times J$ matrix is represented by a sequence of $W$ receptive field, namely the information of $W$ most important nodes is input, while if $J>W$, some less important information will be discarded. Therefore, if a network size disagrees with the input size of a predictor, information loss is more severe in PCR than in LFR-CNN and PATCHY-SAN.

Table \ref{tab:8} shows the prediction errors when all the network sizes are equal to the input size of CNNs, for both training and testing data, namely $H=J=W=800$, with $\langle k\rangle\in[1.5,6]$. Neither upsampling nor downsampling is required for PRC. In this case, all three predictors perform quite well, with very low prediction errors. LFR-CNN outperforms PATCHY-SAN for all 4 networks, and PCR outperforms LFR-CNN for SF network. This suggests that PCR is fragile to the variation of network size. This verifies that LFR makes the prediction performance more robust against network size variation.


\subsection{Run Time Comparison} \label{sub:exp_rt}
\begin{table}[htbp]
	\centering
	\caption{Run time comparison of PCR, PATCHY-SAN, LFR-CNN, and attack simulation (SIM). }
	\begin{tabular}{|c|cc|cc|} \hline
		\begin{tabular}[c]{@{}c@{}}Unit:\\Second\end{tabular}& \multicolumn{2}{c|}{\begin{tabular}[c]{@{}c@{}}Controllability\\ Robustness\end{tabular}} & \multicolumn{2}{c|}{\begin{tabular}[c]{@{}c@{}}Connectivity\\ Robustness\end{tabular}} \\ \hline
		SIM & \multicolumn{2}{c|}{4.7902} & \multicolumn{2}{c|}{1.3704} \\ \hline
		PCR & \multicolumn{2}{c|}{0.0463} & \multicolumn{2}{c|}{0.0477} \\ \hline
		\multirow{2}{*}{PATCHY-SAN} & \multicolumn{1}{c|}{\begin{tabular}[c]{@{}c@{}}LFR\\ 1.1312\end{tabular}} & \begin{tabular}[c]{@{}c@{}}1D-CNN\\ 0.0034\end{tabular} & \multicolumn{1}{c|}{\begin{tabular}[c]{@{}c@{}}LFR\\ 1.1302\end{tabular}} & \begin{tabular}[c]{@{}c@{}}1D-CNN\\ 0.0035\end{tabular} \\ \cline{2-5}
		& \multicolumn{2}{c|}{1.1346} & \multicolumn{2}{c|}{1.1337} \\ \hline
		\multirow{2}{*}{LFR-CNN} & \multicolumn{1}{c|}{\begin{tabular}[c]{@{}c@{}}LFR\\ 1.1320\end{tabular}} & \begin{tabular}[c]{@{}c@{}}CNN\\ 0.0051\end{tabular} & \multicolumn{1}{c|}{\begin{tabular}[c]{@{}c@{}}LFR\\ 1.1300\end{tabular}} & \begin{tabular}[c]{@{}c@{}}CNN\\ 0.0049\end{tabular} \\ \cline{2-5}
		& \multicolumn{2}{c|}{1.1371} & \multicolumn{2}{c|}{1.1349} \\ \hline
	\end{tabular}\label{tab:rt}
\end{table}

Table \ref{tab:rt} shows the run time comparison of PCR, PATCHY-SAN, LFR-CNN, and attack simulation, for both controllability and connectivity robustness predictions. The network size is $N\in[350,650]$; the data are averaged from 100 independent runs. As shown in Table \ref{tab:rt}, the simulation time for controllability robustness is longer than that for connectivity robustness, while for the three predictors, there is no significant difference. It is also notable that PCR is significantly faster than attack simulation, PATCHY-SAN, and LFR-CNN. Running the LFR module is time-consuming, while running the CNN in either PATCHY-SAN or LFR-CNN is faster than PCR due to a simpler structure used. 

Overall, compared to attack simulation, LFR-CNN is able to predict relatively precise controllability and connectivity curves, by saving about $76\%$ and $17\%$ computational time, respectively. In addition, run time for attack simulation increases faster than CNN-based schemes, e.g., with $N\in[700,1300]$, the run time for controllability robustness attack simulation is 41.62 seconds, while it is only 3.67 seconds for LFR-CNN.

\subsection{Compared to Spectral Measures}
\label{sub:exp_spm}

\begin{table*}[htbp]
	\centering
	\caption{Prediction rank errors of the six spectral measures, PCR, PATCHY-SAN, and LFR-CNN. Bold numbers indicate the best performing prediction measures.}
	\begin{tabular}{|c|c|c|c|c|c|c|c|c|c|c|c|}
		\hline
		\begin{tabular}[c]{@{}c@{}}Average\\ Rank Error\end{tabular} & ER & BA & SF & OS & QS & SW-NW & SW-WS & RH & RT & Overall & Rank \\ \hline
		AC & 34.7 & 35.2 & 35.4 & 39.7 & 34.2 & 30.4 & 31.5 & 35.6 & 32.8 & 34.4 & 8 \\ \hline
		EF & \textbf{30.6} & 37.3 & 35.5 & 39.7 & 32.8 & 33.6 & 34.4 & 32.2 & 36.0 & 34.7 & 9 \\ \hline
		NC & 31.8 & 33.6 & 34.2 & 33.7 & 30.9 & 27.9 & 34.0 & 32.1 & 32.7 & 32.3 & 4 \\ \hline
		SG & 31.3 & 33.2 & 31.0 & 34.2 & 34.0 & 29.1 & 32.6 & 33.0 & 35.8 & 32.7 & 6 \\ \hline
		SR & 33.5 & 30.9 & \textbf{29.9} & 33.3 & 33.8 & 31.7 & 30.3 & 34.6 & 33.2 & 32.4 & 5 \\ \hline
		ST & 37.6 & 32.1 & 32.4 & \textbf{30.7} & 34.0 & \textbf{27.2} & 33.0 & 32.0 & \textbf{29.0} & 32.0 & 3 \\ \hline
		PCR & 33.4 & 35.1 & 35.5 & 33.5 & 37.9 & 31.0 & 34.3 & 32.8 & 33.2 & 34.1 & 7 \\ \hline
		PATCHY-SAN & 35.4 & \textbf{28.7} & 30.6 & 31.1 & 32.5 & 28.4 & 30.1 & \textbf{30.3} & 29.6 & \textbf{30.7} & \textbf{1} \\ \hline
		LFR-CNN & 33.5 & 36.7 & 31.7 & 31.6 & \textbf{30.3} & 28.3 & \textbf{29.2} & 30.6 & 29.4 & 31.3 & 2 \\ \hline
	\end{tabular}\label{tab:spt}
\end{table*}

Spectral measures are widely used for estimating network connectivity robustness of undirected networks \cite{Chan2016DMKD}. Table \ref{tab:spt} shows the estimated connectivity robustness ranks of different networks, using three CNN-based predictors and six spectral measures, including algebraic connectivity (AC), effective resistance (EF), natural connectivity (NC), spectral gap (SG), spectral radius (SR), and spanning tree count (ST). Undirected networks with $N\in[350,650]$ and $\langle k\rangle\in[1.5,6]$ are used for comparison. Prediction results are unified by the predicted rank errors of network robustness, calculated by $\xi_{r}=|\hat{rl}-rl|$, where $\hat{rl}$ represents a predicted rank-list and $rl$ is the true rank-list by simulation. For example, given $\hat{rl}=[5,3,1,4,2]$ and $rl=[2,3,1,5,4]$, the rank error is $\xi_{r}=|\hat{rl}-rl|=[3,0,0,1,2]$ and the average rank error is $\bar{\xi}_r=1.2$. As shown in Table \ref{tab:spt}, PATCHY-SAN and LFR-CNN obtain the best two average rank errors, while PCR does not perform well due to a larger variation of network size and average degree.

\section{Conclusion}\label{sec:end}

In this paper, a learning feature representation-based convolutional neural network, namely LFR-CNN, is developed for network robustness performance prediction, including both connectivity robustness and controllability robustness. Conventionally, network robustness is evaluated by time-consuming attack simulations, from which a sequence of network connectivity or controllability values are collected and used to measure the remaining network after a sequence of destructive attacks (here, node-removal attacks). LFR-CNN is designed to gain a balance between PCR and PATCHY-SAN, in terms of both input size and internal parameters. The LFR module not only compresses the raw higher-dimensional adjacency matrix to a lower-dimensional representation, but also extends the capability of LFR-CNN to process complex network data with a wide-ranged variation of network size and average degree.

Extensive numerical experiments are performed using both synthetic and real-world networks, including directed and undirected networks, and then analyzed and compared, revealing clearly the pros and cons of several typical and comparable schemes and measures. Specifically, the good performance of LFR-CNN in predicting both connectivity robustness and controllability robustness is verified by comparing with other two state-of-the-art network robustness predictors, namely PCR and PATCHY-SAN. LFR-CNN is much less sensitive than PCR to the network size variation. Although LFR-CNN requires a relatively long run time for feature learning, it can still achieve accurate prediction faster than the conventional attack simulations. Meanwhile, LFR-CNN not only can accurately predict the connectivity and controllability robustness curves of various complex networks under different types of attacks, but also serves as an excellent indicator for the connectivity robustness, better than spectral measures.

The present study, after all, makes the current investigation of network controllability and connectivity robustness more subtle and complete. Yet, it should be noted that the correlation between controllability robustness and spectral measures has not been investigated, leaving a good but challenging topic for future research.

\bibliographystyle{IEEEtran}
\bibliography{ref}
	
\end{document}